\documentclass[12pt,aps,prd,showpacs,amsmath,amssymb]{revtex4}
\usepackage{graphicx}
\usepackage{epstopdf}
\usepackage{multirow}
\usepackage{subfigure}
\usepackage{appendix}
\input epsf
\begin{document}
\title{$\bar{D}^{(*)}_{s}D^{(*)}$ molecular state with $J^{P}=1^{+}$}
\author{Yong-Jiang Xu$^{1}$\footnote{xuyongjiang13@nudt.edu.cn}, Yong-Lu Liu$^1$, Chun-Yu Cui$^2$, and Ming-Qiu Huang$^{1,3}$\footnote{Corresponding author. mqhuang@nudt.edu.cn}}
\affiliation{$^1$Department of Physics, College of Liberal Arts and Sciences, National University of Defense Technology , Changsha, 410073, Hunan, China}
\affiliation{$^2$Department of Physics, College of Basic Medical Science, Army Medical University, Chongqing, 400038, China}
\affiliation{$^3$Synergetic Innovation Center for Quantum Effects and Applications, Hunan Normal University, Changsha,  410081, Hunan, China}
\date{}
\begin{abstract}
In this paper, we construct $\bar{D}^{(*)}_{s}D^{(*)}$-molecule-type interpolating currents $J_{(\pm)\mu}(x)$ with $J^{P}=1^{+}$, calculate the corresponding mass and magnetic moment using the QCD sum rule method and its extension in the weak electromagnetic field, and study the processes of $Z_{(\pm)cs}$ to $\eta_{c}K^{*}$, $J/\psi K$, $\bar{D}D^{*}_{s}$, and $\bar{D}^{*}D_{s}$ via three-point sum rules. The numerical values are $m_{Z_{(\pm)cs}}=3.99^{+0.17}_{-0.14}~\mbox{GeV}$, and $\lambda_{Z_{(\pm)cs}}=2.07^{+0.28}_{-0.16}\times10^{-2}~\mbox{GeV}^{5}$, $\mu_{Z_{(\pm)cs}}=0.18^{+0.16}_{-0.09}~\mu_{N}$ with $\mu_{N}$ the nucleon magneton, $\Gamma_{Z_{(+)cs}}=17.47^{+12.70}_{-8.08}$, and $\Gamma_{Z_{(-)cs}}=13.86^{+10.37}_{-6.51}$. The masses are in agreement with the recently measured value of $Z_{cs}(3985)$ by the BESIII Collaboration, $m^{exp}_{Z_{cs}}=(3982.5^{+1.8}_{-2.6}\pm2.1)~\mbox{MeV}$. The widths are compatible with the experimental value, $\Gamma^{exp}_{Z_{cs}}=(12.8^{+5.3}_{-4.4}\pm3.0)~\mbox{MeV}$. The magnetic moment and the various decay modes can help us to determine the inner structure of $Z_{cs}(3985)$ when being confronted with experimental data in the future.
\end{abstract}
 \maketitle

\section{Introduction}\label{sec1}

Since the observation of $X(3872)$ \cite{belle1}, there have been plenty of $XYZ$ states, the so-called exotic hadron states, reported experimentally, and many theoretical studies have been done to understand their various properties \cite{PDG,H.X.Chen1,H.X.Chen2,H.X.Chen3}. In particular, the exotic hadron states with strange flavor have attracted theorists' attention, and different methods and models have been used to study them \cite{strange1,strange2,strange3,strange4,strange5,strange6,strange7,strange8,strange9,strange10,strange11,strange12}.

Very recently, the BESIII Collaboration reported a new structure $Z_{cs}(3985)$ in the $K^{+}$ recoil-mass spectrum near the $D^{-}_{s}D^{*0}$/$D^{*-}_{s}D^{0}$ mass thresholds in the processes of $e^{+}e^{-}\rightarrow K^{+}(D^{-}_{s}D^{*0}+D^{*-}_{s}D^{0})$ at $\sqrt{s}=4.681~\mbox{GeV}$ \cite{BES}. Its pole mass and width are measured as $(3982.5^{+1.8}_{-2.6}\pm2.1)~\mbox{MeV}$ and $(12.8^{+5.3}_{-4.4}\pm3.0)~\mbox{MeV}$, respectively. It can decay into $D^{-}_{s}D^{*0}$ and $D^{*-}_{s}D^{0}$ in S wave, its spin-parity is assumed to favor $J^{P}=1^{+}$ and the quark content as $\bar{c}cs\bar{u}$ \cite{BES}. It will be the first candidate of the hidden-charm four-quark state with strangeness. Inspired by this observation, many theoretical works have appeared concerning the new observed state's mass, production and decay \cite{strange13,strange14,strange15,strange16,strange17,strange18,strange19,strange20,strange21,strange22,strange23,strange24,strange25,strange26,strange27,strange28}.

In the present work, we construct $\bar{D}^{(*)}_{s}D^{(*)}$-molecule-type interpolating currents $J_{(\pm)\mu}(x)$ with $J^{P}=1^{+}$, calculate the corresponding mass and magnetic moment using the QCD sum rule method and its extension in the weak electromagnetic field, and study their decay properties via three-point sum rule. The QCD sum rule method \cite{SVZ1,SVZ2} is a nonperturbative analytic formalism firmly entrenched in QCD with minimal modeling and has been successfully applied in almost every aspect of strong interaction physics. Its extension in the weak electromagnetic field can be used to calculate the magnetic moment of ground hadron states \cite{Balitsky,Ioffe1,Ioffe2,octet1,octet2,octet3,octet4,decuplet1,decuplet2,decuplet3,decuplet4,rho,wangzhigang,
xuyongjiang1,xuyongjiang2}. The electromagnetic multipole moments of the hadron encode the spatial distributions of charge and magnetization in the hadron and provide important information about the quark configurations of the hadron and the underlying dynamics. So it is interesting to study the electromagnetic multipole moments of the hadron.

The rest of the paper is organized as follows. In Sec.\ref{sec2}, the relevant sum rules are derived. Section~\ref{sec3} is devoted to the numerical analysis, and a short summary is given in Sec.\ref{sec4}. In Appendix \ref{appendix}, the spectral densities are shown.

\section{The derivation of the sum rules}\label{sec2}

\subsection{Mass and magnetic moment}
First, we write down the molecule-type interpolating currents with $J^{P}=1^{+}$:
\begin{equation}\label{Zcs interpolating current}
J_{(\pm)\mu}(x)=\frac{1}{\sqrt{2}}\{[\bar{s}(x)i\gamma^{5}c(x)][\bar{c}(x)\gamma_{\mu}u(x)]\pm[\bar{s}(x)\gamma_{\mu}c(x)][\bar{c}(x)i\gamma^{5}u(x)]\},
\end{equation}
which can couple to the $\bar{D}^{(*)}_{s}D^{(*)}$ molecular state (labeled as $Z_{(\pm)cs}$), and the coupling strength can be parameterized as follows:
\begin{equation}\label{poleresidue}
\langle0|J_{(\pm)\mu}(0)|Z_{(\pm)cs}(p)\rangle=\lambda_{Z_{(\pm)cs}}\epsilon_{(\pm)\mu}(p)
\end{equation}
with $\lambda_{Z_{(\pm)cs}}$ and $\epsilon_{(\pm)\mu}(p)$ being the pole residue and polarization vector of $Z_{(\pm)cs}$ state, respectively.

The charge, magnetic, and quadrupole form factors of the $Z_{(\pm)cs}$ state are related to three functions- $G_{1}(Q^{2})$, $G_{2}(Q^{2})$, and $G_{3}(Q^{2})$:
\begin{eqnarray}
&&G_{(\pm)C}(Q^{2})=G_{(\pm)1}(Q^{2})+\frac{2}{3}\eta G_{(\pm)Q}(Q^{2}),\nonumber\\
&&G_{(\pm)M}(Q^{2})=-G_{(\pm)2}(Q^{2}),\nonumber\\
&&G_{(\pm)Q}(Q^{2})=G_{(\pm)1}(Q^{2})+G_{(\pm)2}(Q^{2})+(1+\eta)G_{(\pm)3}(Q^{2}),
\end{eqnarray}
with $\eta=\frac{Q^{2}}{4m^{2}_{Z_{(\pm)cs}}}$ and the functions $G_{(\pm)1}(Q^{2})$, $G_{(\pm)2}(Q^{2})$, and $G_{(\pm)3}(Q^{2})$ defined as
\begin{eqnarray}
\langle Z_{(\pm)cs}(p)|j^{em}_{\alpha}(0)|Z_{(\pm)cs}(p^{\prime})\rangle=&&G_{(\pm)1}(Q^{2})\epsilon^{*}_{(\pm)}(p)\cdot\epsilon_{\pm}(p^{\prime})
(p+p^{\prime})_{\alpha}+G_{(\pm)2}(Q^{2})[\epsilon_{(\pm)\alpha}(p^{\prime})\epsilon^{*}_{(\pm)}(p)\cdot q\nonumber\\&&-\epsilon^{*}_{(\pm)\alpha}(p)\epsilon_{(\pm)}(p^{\prime})\cdot q]
-\frac{G_{(\pm)3}(Q^{2})}{2m^{2}_{Z_{c}}}\epsilon^{*}_{(\pm)}(p)\cdot q\epsilon_{(\pm)}(p^{\prime})\cdot q(p+p^{\prime})_{\alpha},
\end{eqnarray}
with $q=p^{\prime}-p$ and $Q^{2}=-q^{2}$. At zero momentum transfer, these form factors are proportional to the usual static quantities of the charge $e$, magnetic moment $\mu_{Z_{(\pm)cs}}$, and quadrupole moment $Q_{(\pm)1}$:
\begin{eqnarray}
&&eG_{(\pm)C}(0)=e,\nonumber\\
&&eG_{(\pm)M}(0)=2m_{Z_{(\pm)cs}}\mu_{Z_{(\pm)cs}},\nonumber\\
&&eG_{(\pm)Q}(0)=m^{2}_{Z_{(\pm)cs}}Q_{(\pm)1}.
\end{eqnarray}

To derive the needed sum rules, we begin with the time-ordered correlation function in the QCD vacuum in the presence of a constant background electromagnetic field $F_{\mu\nu}$:
\begin{equation}\label{2-point correlator}
\Pi_{(\pm)\mu\nu}(p)=i\int dx^{4}e^{ipx}\langle0\mid\textsl{T}[J_{(\pm)\mu}(x)J^{\dagger}_{(\pm)\nu}(0)]\mid0\rangle_{F}
       =\Pi^{(0)}_{(\pm)\mu\nu}(p)+\Pi^{(1)}_{(\pm)\mu\nu\alpha\beta}(p)F^{\alpha\beta}+\cdots,
\end{equation}
where $J_{(\pm)\mu}(x)$ is the interpolating current of $Z_{(\pm)cs}$ state (\ref{Zcs interpolating current}). The $\Pi^{(0)}_{(\pm)\mu\nu}(p)$ term is the correlation function without an external electromagnetic field, and gives rise to the mass and pole residue of $Z_{(\pm)cs}$. The magnetic moment will be extracted from the linear response term $\Pi^{(1)}_{(\pm)\mu\nu\alpha\beta}(p)F^{\alpha\beta}$.

Following the method stated in Ref. \cite{xuyongjiang1,xuyongjiang2}, we express physically the correlation function (\ref{2-point correlator}) as
\begin{eqnarray}\label{hadronic side}
\Pi^{had}_{(\pm)\mu\nu}(p)=&&\frac{\lambda^{2}_{Z_{(\pm)cs}}}{m^{2}_{Z_{(\pm)cs}}-p^{2}}(-g_{\mu\nu}+\frac{p_{\mu}p_{\nu}}{p^{2}})
\nonumber\\&&-i\frac{\lambda^{2}_{Z_{(\pm)cs}}G_{(\pm)2}(0)}{(p^{2}-m^{2}_{Z_{(\pm)cs}})^{2}}F_{\mu\nu}+i\frac{a}{m^{2}_{Z_{(\pm)cs}}-p^{2}}F_{\mu\nu}+\cdots,
\end{eqnarray}
where the constant $a$ parametrizes the contributions from the pole-continuum transitions.

In Eq. (\ref{2-point correlator}), we substitute $J_{(\pm)\mu}(x)$ with Eq. (\ref{Zcs interpolating current}), contract the relevant quark fields via Wick's theorem, and obtain
\begin{eqnarray}
 \Pi^{OPE}_{(\pm)\mu\nu}(p)=\frac{i}{2}\int d^{4}xe^{ipx}&&(Tr[(i\gamma_{5})S^{(s)}_{ca}(-x)(i\gamma_{5})S^{(c)}_{ac}(x)]Tr[\gamma_{\mu}S^{(u)}_{bd}(x)\gamma_{\nu}S^{(c)}_{db}(-x)]\nonumber\\
 &&\pm Tr[(i\gamma_{5})S^{(c)}_{db}(-x)\gamma_{\mu}S^{(u)}_{bd}(x)]Tr[(i\gamma_{5})S^{(c)}_{ac}(x)\gamma_{\nu}S^{(s)}_{ca}(-x)]\nonumber\\
 &&\pm Tr[(i\gamma_{5})S^{(s)}_{ca}(-x)\gamma_{\mu}S^{(c)}_{ac}(x)]Tr[(i\gamma_{5})S^{(u)}_{bd}(x)\gamma_{\nu}S^{(c)}_{db}(-x)]\nonumber\\
 &&+Tr[(i\gamma_{5})S^{(u)}_{bd}(x)(i\gamma_{5})S^{(c)}_{db}(-x)]Tr[\gamma_{\mu}S^{(c)}_{ac}(x)\gamma_{\nu}S^{(s)}_{ca}(-x)]),
\end{eqnarray}
where $S^{(c)}(x)=\langle 0|T[c(x)\bar{c}(0)]|0\rangle$ and $S^{(q)}(x)=\langle 0|T[q(x)\bar{q}(0)]|0\rangle, q=u, s$ are the full charm- and up (strange)-quark propagators, respectively, whose expressions are given in Appendix \ref{appendix1}, $Tr$ denotes the trace of the Dirac spinor indices, and $a$, $b$, $c$, and $d$ are color indices. Through dispersion relation, $\Pi^{OPE}_{(\pm)\mu\nu}(p)$ can be written as
 \begin{eqnarray}\label{QCD side}
 \Pi^{OPE}_{(\pm)\mu\nu}(p)=&&\int^{\infty}_{(2m_{c}+m_{s})^{2}}ds\frac{\rho^{(0)}_{(\pm)}(s)}{s-p^{2}}(-g_{\mu\nu}+\frac{p_{\mu}p_{\nu}}{p^{2}})
 +\int^{\infty}_{(2m_{c}+m_{s})^{2}}ds\frac{\rho^{(1)}_{(\pm)}(s)}{s-p^{2}}(iF_{\mu\nu})\nonumber\\&&+\mbox{other Lorentz structures},
 \end{eqnarray}
 where $\rho^{i}_{(\pm)}(s)=\frac{1}{\pi}\mbox{Im}\Pi^{OPE}_{(\pm)i}(s),~i=0,1$ are the spectral densities, and $m_{c}$ and $m_{s}$ are the masses of the charm and strange quark, respectively. We find that $\rho^{i}_{(+)}(s)=\rho^{i}_{(-)}(s),~i=0,1$, and do not distinguish the subscripts $\pm$ in the rest of this subsection. The expressions of $\rho^{i}(s),~i=0,1$, are given in Appendix \ref{appendix}.

 Finally, matching the phenomenological side (\ref{hadronic side}) and the QCD representation (\ref{QCD side}), we obtain
 \begin{equation}
 \frac{\lambda^{2}_{Z_{cs}}}{m^{2}_{Z_{cs}}-p^2}+\cdots=\int^{\infty}_{(2m_{c}+m_{s})^{2}}ds\frac{\rho^{(0)}(s)}{s-p^2},
 \end{equation}
 for the Lorentz structure $(-g_{\mu\nu}+\frac{p_{\mu}p_{\nu}}{p^{2}})$, and
 \begin{equation}
 \frac{\lambda^{2}_{Z_{cs}}G_{M}(0)}{(m^{2}_{Z_{cs}}-p^{2})^2}+\frac{a}{m^{2}_{Z_{cs}}-p^2}+\cdots=\int^{\infty}_{(2m_{c}+m_{s})^{2}}ds\frac{\rho^{(1)}(s)}{s-p^2},
 \end{equation}
 for the Lorentz structure $iF_{\mu\nu}$.

 According to quark-hadron duality, the excited and continuum states' spectral density can be approximated by the QCD spectral density above some effective threshold $s^{0}_{Z_{cs}}$, whose value will be determined in Sec.\ref{sec3}:
 \begin{eqnarray}
 &&\frac{\lambda^{2}_{Z_{cs}}}{m^{2}_{Z_{cs}}-p^2}+\int^{\infty}_{s^{0}_{Z_{cs}}}ds\frac{\rho^{(0)}(s)}{s-p^2}=\int^{\infty}_{(2m_{c}+m_{s})^{2}}ds\frac{\rho^{(0)}(s)}{s-p^2},\nonumber\\
 &&\frac{\lambda^{2}_{Z_{cs}}G_{M}(0)}{(m^{2}_{Z_{cs}}-p^{2})^2}+\frac{a}{m^{2}_{Z_{cs}}-p^2}+\int^{\infty}_{s^{0}_{Z_{cs}}}ds\frac{\rho^{(1)}(s)}{s-p^2}=\int^{\infty}_{(2m_{c}+m_{s})^{2}}ds\frac{\rho^{(1)}(s)}{s-p^2}.
 \end{eqnarray}
 Subtracting the contributions of the excited and continuum states, one gets
 \begin{eqnarray}
 &&\frac{\lambda^{2}_{Z_{cs}}}{m^{2}_{Z_{cs}}-p^2}=\int^{s^{0}_{Z_{cs}}}_{(2m_{c}+m_{s})^{2}}ds\frac{\rho^{(0)}(s)}{s-p^2},\nonumber\\
 &&\frac{\lambda^{2}_{Z_{cs}}G_{M}(0)}{(m^{2}_{Z_{cs}}-p^{2})^2}+\frac{a}{m^{2}_{Z_{cs}}-p^2}=\int^{s^{0}_{Z_{cs}}}_{(2m_{c}+m_{s})^{2}}ds\frac{\rho^{(1)}(s)}{s-p^2}.
 \end{eqnarray}

 In order to improve the convergence of the operator production expansion (OPE) series and suppress the contributions from the excited and continuum states, it is necessary to make a Borel transform. As a result, we have
 \begin{eqnarray}\label{2-point sum rule}
 &&\lambda^{2}_{Z_{cs}}e^{-m^{2}_{Z_{cs}}/M^{2}_{B}}=\int^{s^{0}_{Z_{cs}}}_{(2m_{c}+m_{s})^{2}}ds\rho^{(0)}(s)e^{-s/M^{2}_{B}},\nonumber\\
 &&\lambda^{2}_{Z_{cs}}(\frac{G_{M}(0)}{M^{2}_{B}}+A)e^{-m^{2}_{Z_{cs}}/M^{2}_{B}}=\int^{s^{0}_{Z_{cs}}}_{(2m_{c}+m_{s})^{2}}ds\rho^{(1)}(s)e^{-s/M^{2}_{B}}.
 \end{eqnarray}
 where $M^{2}_{B}$ is the Borel parameter and $A=\frac{a}{\lambda^{2}_{Z_{cs}}}$. Taking the derivative of the first equation in (\ref{2-point sum rule}) with respect to $-\frac{1}{M^{2}_{B}}$ and dividing it by the original expression, one has
 \begin{equation}\label{mass}
 m^{2}_{Z_{cs}}=\frac{\frac{d}{d(-\frac{1}{M^{2}_{B}})}\int^{s^{0}_{Z_{cs}}}_{(2m_{c}+m_{s})^{2}}ds\rho^{(0)}(s)e^{-\frac{s}{M^{2}_{B}}}}{\int^{s^{0}_{Z_{cs}}}_{(2m_{c}+m_{s})^{2}}ds\rho^{(0)}(s)e^{-\frac{s}{M^{2}_{B}}}}.
 \end{equation}

\subsection{Strong decay form factors}

In this subsection, we calculate the strong decay form factors of $Z_{(\pm)cs}$ to $\eta_{c}K^{*}$, $J/\psi K$, $\bar{D}D^{*}_{s}$, and $\bar{D}^{*}D_{s}$. To this end, we start with the following three-point functions:
\begin{eqnarray}
&&\Gamma^{1}_{(\pm)\mu\nu}(p,p_{1},p_{2})=i^{2}\int d^{4}xd^{4}ye^{i(p_{1}x+p_{2}y)}\langle0|T[J^{\eta_{c}}(x)J^{K^{*}}_{\mu}(y)J^{\dagger}_{(\pm)\nu}(0)]|0\rangle,
\nonumber\\&&\Gamma^{2}_{(\pm)\mu\nu}(p,p_{1},p_{2})=i^{2}\int d^{4}xd^{4}ye^{i(p_{1}x+p_{2}y)}\langle0|T[J^{J/\psi}_{\mu}(x)J^{K}(y)J^{\dagger}_{(\pm)\nu}(0)]|0\rangle,
\nonumber\\&&\Gamma^{3}_{(\pm)\mu\nu}(p,p_{1},p_{2})=i^{2}\int d^{4}xd^{4}ye^{i(p_{1}x+p_{2}y)}\langle0|T[J^{\bar{D}}(x)J^{D^{*}_{s}}_{\mu}(y)J^{\dagger}_{(\pm)\nu}(0)]|0\rangle,
\nonumber\\&&\Gamma^{4}_{(\pm)\mu\nu}(p,p_{1},p_{2})=i^{2}\int d^{4}xd^{4}ye^{i(p_{1}x+p_{2}y)}\langle0|T[J^{\bar{D}^{*}}_{\mu}(x)J^{D_{s}}(y)J^{\dagger}_{(\pm)\nu}(0)]|0\rangle,
\end{eqnarray}
where $J_{(\pm)\nu}(x)$ is the interpolating current of $Z_{(\pm)cs}$, and $J^{\eta_{c}}(x)$, $J^{K^{*}}_{\mu}(x)$, $J^{J/\psi}_{\mu}(x)$, $J^{K}(x)$, $J^{\bar{D}}(x)$, $J^{D^{*}_{s}}_{\mu}(x)$, $J^{\bar{D}^{*}}_{\mu}(x)$, and $J^{D_{s}}(x)$ are the interpolating currents of $\eta_{c}$, $K^{*}$, $J/\psi$, $K$, $\bar{D}$, $D^{*}_{s}$, $\bar{D}^{*}$, and $D_{s}$, respectively, being given by
\begin{eqnarray}
&&J^{\eta_{c}}(x)=\bar{c}(x)i\gamma_{5}c(x),\nonumber\\
&&J^{K^{*}}_{\mu}(x)=\bar{s}(s)\gamma_{\mu}u(x),\nonumber\\
&&J^{J/\psi}_{\mu}(x)=\bar{c}(x)\gamma_{\mu}c(x),\nonumber\\
&&J^{K}(x)=\bar{s}(x)i\gamma_{5}u(x),\nonumber\\
&&J^{\bar{D}}(x)=\bar{c}(x)i\gamma_{5}u(x),\nonumber\\
&&J^{D^{*}_{s}}_{\mu}(x)=\bar{s}(s)\gamma_{\mu}c(x),\nonumber\\
&&J^{\bar{D}^{*}}_{\mu}(x)=\bar{c}(x)\gamma_{\mu}u(x),\nonumber\\
&&J^{D_{s}}(x)=\bar{s}(x)i\gamma_{5}c(x).
\end{eqnarray}

We will take $\Gamma^{1}_{(+)\mu\nu}(p,p_{1},p_{2})$ as an example to illustrate the steps involved in our calculation. Inserting complete sets of hadronic states into the three-point correlation function, we have
\begin{eqnarray}\label{hadrnic side 1}
\Gamma^{1phy}_{(+)\mu\nu}=&&[g_{1(+)}\frac{\lambda_{Z_{cs}}f_{\eta_{c}}f_{K^{*}}m^{2}_{\eta_{c}}m_{K^{*}}/(2m_{c})}
{(m^{2}_{\eta_{c}}-p^{2}_{1})(m^{2}_{K^{*}}-p^{2}_{2})(m^{2}_{Z_{cs}}-p^{2})}
+\frac{A}{(m^{2}_{\eta_{c}}-p^{2}_{1})(m^{2}_{K^{*}}-p^{2}_{2})}]g_{\mu\nu}\nonumber\\
&&+\mbox{other Lorentz structures}+\cdots,
\end{eqnarray}
where we make use of Eq.(\ref{poleresidue}) and the following matrix elements:
\begin{eqnarray}
&&\langle0|J^{\eta_{c}}(0)|\eta_{c}(p_{1})\rangle=\frac{f_{\eta_{c}}m^{2}_{\eta_{c}}}{2m_{c}},\nonumber\\
&&\langle0|J^{K^{*}}_{\mu}(0)|K^{*}(p_{2})\rangle=f_{K^{*}}m_{K^{*}}\epsilon_{\mu}(p_{2}),\nonumber\\
&&\langle \eta_{c}(p_{1})K^{*}(p_{2})|Z_{(+)cs}(p)\rangle=g_{1(+)}\epsilon^{\ast}(p_{2})\cdot\epsilon(p).
\end{eqnarray}
The constant $A$ represents the transition between ground state and excited states, which cannot be ignored.

On the other hand, we can calculate $\Gamma^{1}_{(+)\mu\nu}(p,p_{1},p_{2})$ theoretically. Substituting the interpolating currents in $\Gamma^{1}_{(+)\mu\nu}(p,p_{1},p_{2})$ with their explicit expressions and contracting the quark fields via Wick's theorem, we obtain
\begin{eqnarray}
\Gamma^{1}_{(+)\mu\nu}(p,p_{1},p_{2})=&&-\frac{1}{\sqrt{2}}\int\frac{d^{4}k}{(2\pi)^{4}}\frac{d^{4}q}{(2\pi)^{4}}
d^{4}xd^{4}y\{Tr[i\gamma_{5}S^{(c)}_{ac}(k)i\gamma_{5}S^{(s)}_{cb}(-y)\gamma_{\mu}S^{(u)}_{bd}(y)\gamma_{\nu}S^{(c)}_{da}(-q)]
\nonumber\\&&+ Tr[i\gamma_{5}S^{(c)}_{ac}(k)\gamma_{\nu}S^{(s)}_{cb}(-y)\gamma_{\mu}S^{(u)}_{bd}(y)i\gamma_{5}S^{(c)}_{da}(-q)]\}e^{i(p_{1}-k-q)x+ip_{2}y}.
\end{eqnarray}
Substituting the quark propagators with their explicit expressions and carrying out the integrals, we obtain the theoretical side of the correlation function:
\begin{equation}\label{QCD side 1}
\Gamma^{1OPE}_{(+)\mu\nu}(p,p_{1},p_{2})=\Gamma^{1OPE}_{(+)}(p^{2},p^{2}_{1},p^{2}_{2})g_{\mu\nu}+\mbox{other Lorentz structures},
\end{equation}
where the coefficient $\Gamma^{1OPE}_{(+)}(p^{2},p^{2}_{1},p^{2}_{2})$ can be written as
\begin{equation}\label{QCD side 2}
\Gamma^{1OPE}_{(+)}(p^{2},p^{2}_{1},p^{2}_{2})=\int^{\infty}_{4m^{2}_{c}}dt_{1}\int^{\infty}_{m^{2}_{s}}dt_{2}
\frac{\rho^{1}_{3(+)}(p^{2},t_{1},t_{2})}{(t_{1}-p^{2}_{1})(t_{2}-p^{2}_{2})},
\end{equation}
with $\rho^{1}_{3(+)}(p^{2},t_{1},t_{2})$ being the QCD special density. The explicit expressions of $\rho^{1}_{3(+)}(p^{2},t_{1},t_{2})$ are given in Appendix \ref{appendix}.

It is time to match the physical representation (\ref{hadrnic side 1}) and theoretical representation (\ref{QCD side 1}) of $\Gamma^{1}_{(+)\mu\nu}(p,p_{1},p_{2})$. For the Lorentz structure $g_{\mu\nu}$, we have, by using the quark-hadron duality,
\begin{eqnarray}
&&g_{1(+)}\frac{\lambda_{Z_{cs}}f_{\eta_{c}}f_{K^{*}}m^{2}_{\eta_{c}}m_{K^{*}}/(2m_{c})}
{(m^{2}_{\eta_{c}}-p^{2}_{1})(m^{2}_{K^{*}}-p^{2}_{2})(m^{2}_{Z_{cs}}-p^{2})}+\frac{A}{(m^{2}_{\eta_{c}}-p^{2}_{1})(m^{2}_{K^{*}}-p^{2}_{2})}
\nonumber\\&&=\int^{t^{0}_{1}}_{4m^{2}_{c}}dt_{1}\int^{t^{0}_{2}}_{m^{2}_{s}}dt_{2}
\frac{\rho^{1}_{3(+)}(p^{2},t_{1},t_{2})}{(t_{1}-p^{2}_{1})(t_{2}-p^{2}_{2})},
\end{eqnarray}
where $t^{0}_{1}$ and $t^{0}_{2}$ are the threshold parameters in the channels of $\eta_{c}$ and $K^{*}$, respectively.

Setting $p^{2}=p^{2}_{1}$ and performing Borel transforms $p^{2}_{1}\rightarrow M^{2}_{B1}$ and $p^{2}_{2}\rightarrow M^{2}_{B2}$, one gets
\begin{eqnarray}
&&g_{1(+)}\frac{\lambda_{Z_{cs}}f_{\eta_{c}}f_{K^{*}}m^{2}_{\eta_{c}}m_{K^{*}}}{2m_{c}(m^{2}_{Z_{cs}}-m^{2}_{\eta_{c}})}
(e^{-m^{2}_{\eta_{c}}/M^{2}_{B1}}-e^{-m^{2}_{Z_{cs}}/M^{2}_{B1}})e^{-m^{2}_{K^{*}}/M^{2}_{B2}}
+Ae^{-m^{2}_{\eta_{c}}/M^{2}_{B1}}e^{-m^{2}_{K^{*}}/M^{2}_{B2}}\nonumber\\&&
=\int^{t^{0}_{1}}_{4m^{2}_{c}}dt_{1}\int^{t^{0}_{2}}_{m^{2}_{s}}dt_{2}e^{-t_{1}/M^{2}_{B1}}e^{-t_{2}/M^{2}_{B2}}\rho^{1}_{3(+)}(t_{1},t_{1},t_{2}).
\end{eqnarray}
Acting on the above equation with operator $m^{2}_{\eta_{c}}-\frac{d}{d(-1/M^{2}_{B1})}$, we obtain
\begin{eqnarray}
&&g_{1(+)}\frac{\lambda_{Z_{cs}}f_{\eta_{c}}f_{K^{*}}m^{2}_{\eta_{c}}m_{K^{*}}}{2m_{c}}e^{-m^{2}_{Z_{cs}}/M^{2}_{B1}}e^{-m^{2}_{K^{*}}/M^{2}_{B2}}
\nonumber\\&&=[m^{2}_{\eta_{c}}-\frac{d}{d(-\frac{1}{M^{2}_{B1}})}]\int^{t^{0}_{1}}_{4m^{2}_{c}}dt_{1}\int^{t^{0}_{2}}_{m^{2}_{s}}dt_{2}e^{-t_{1}/M^{2}_{B1}}e^{-t_{2}/M^{2}_{B2}}\rho^{1}_{3(+)}(t_{1},t_{1},t_{2}).
\end{eqnarray}

We can study similarly other three-point correlation functions, and the corresponding spectral densities are given in Appendix \ref{appendix}.

\section{Numerical analysis}\label{sec3}

The input parameters needed in numerical analysis are $\langle\bar{q}q\rangle=-(0.24\pm0.01)^{3}~\mbox{GeV}^{3}$, $\langle\bar{s}s\rangle=(0.8\pm0.1)\langle0|\bar{q}q|0\rangle$, $\langle g_{s}\bar{q}\sigma Gq\rangle=(0.8\pm0.1)\langle0|\bar{q}q|0\rangle~\mbox{GeV}^{2}$, $\langle g_{s}\bar{s}\sigma Gs\rangle=(0.8\pm0.1)\langle0|\bar{s}s|0\rangle \mbox{GeV}^{2}$, $\langle g^{2}_{s}GG\rangle=0.88\pm0.25~\mbox{GeV}^{4}$, $m_{c}=(1.275\pm0.025)~\mbox{GeV}$, $m_{s}=(0.095\pm0.005)~\mbox{GeV}$, $m_{\eta_{c}}=2.98~\mbox{GeV}$, $m_{K^{*}}=0.89~\mbox{GeV}$, $m_{J/\psi}=3.07~\mbox{GeV}$, $m_{K}=0.49~\mbox{GeV}$, $m_{D}=1.86~\mbox{GeV}$, $m_{D^{*}_{s}}=2.11~\mbox{GeV}$, $m_{D^{*}}=2.01~\mbox{GeV}$, $m_{D_{s}}=1.97~\mbox{GeV}$, $f_{\eta_{c}}=0.35~\mbox{GeV}$, $f_{K^{*}}=0.22~\mbox{GeV}$, $f_{J/\psi}=0.41~\mbox{GeV}$, $f_{K}=0.16~\mbox{GeV}$, $f_{D}=0.18~\mbox{GeV}$, $f_{D^{*}_{s}}=0.33~\mbox{GeV}$, $f_{D^{*}}=0.24~\mbox{GeV}$, and $f_{D_{s}}=0.24~\mbox{GeV}$, which can be found in Ref. \cite{strange15}. For the vacuum susceptibilities $\chi$, $\kappa$, and $\xi$, we take the values $\chi=-(3.15\pm0.30)~\mbox{GeV}^{-2}$, $\kappa=-0.2$, and $\xi=0.4$ determined in the detailed QCD sum rules analysis of the photon light-cone distribution amplitudes \cite{P.Ball}. Besides these parameters, we should determine the working intervals of the threshold parameters and the Borel mass in which the mass, the pole residue, the magnetic moment, and strong decay form factors vary weakly. The continuum threshold is related to the square of the first excited states having the same quantum number as the interpolating field, while the Borel parameter is determined by demanding that both the contributions of the higher states and continuum are sufficiently suppressed and the contributions coming from higher-dimensional operators are small.

We define two quantities--the ratio of the pole contribution to the total contribution (RP) and the ratio of the highest-dimensional term in the OPE series to the total OPE series (RH) as follows:
\begin{eqnarray}
&&RP_{i}\equiv\frac{\int^{s^{0}_{Z_{cs}}}_{(2m_{c}+m_{s})^{2}}ds\rho^{(i)}(s)e^{-\frac{s}{M^{2}_{B}}}}{\int^{\infty}_{(2m_{c}+m_{s})^{2}}ds\rho^{(i)}(s)e^{-\frac{s}{M^{2}_{B}}}},
\nonumber\\&&RH_{i}\equiv\frac{\int^{s^{0}_{Z_{cs}}}_{(2m_{c}+m_{s})^{2}}ds\rho^{(d=8)}_{i}(s)e^{-\frac{s}{M^{2}_{B}}}}{\int^{s^{0}_{Z_{cs}}}_{(2m_{c}+m_{s})^{2}}ds\rho^{(i)}(s)e^{-\frac{s}{M^{2}_{B}}}},
\end{eqnarray}
with $i=0,1$. Similar quantities can be defined for three-point correlation functions.

\begin{figure}[htb]
\subfigure[]{
\includegraphics[width=7cm]{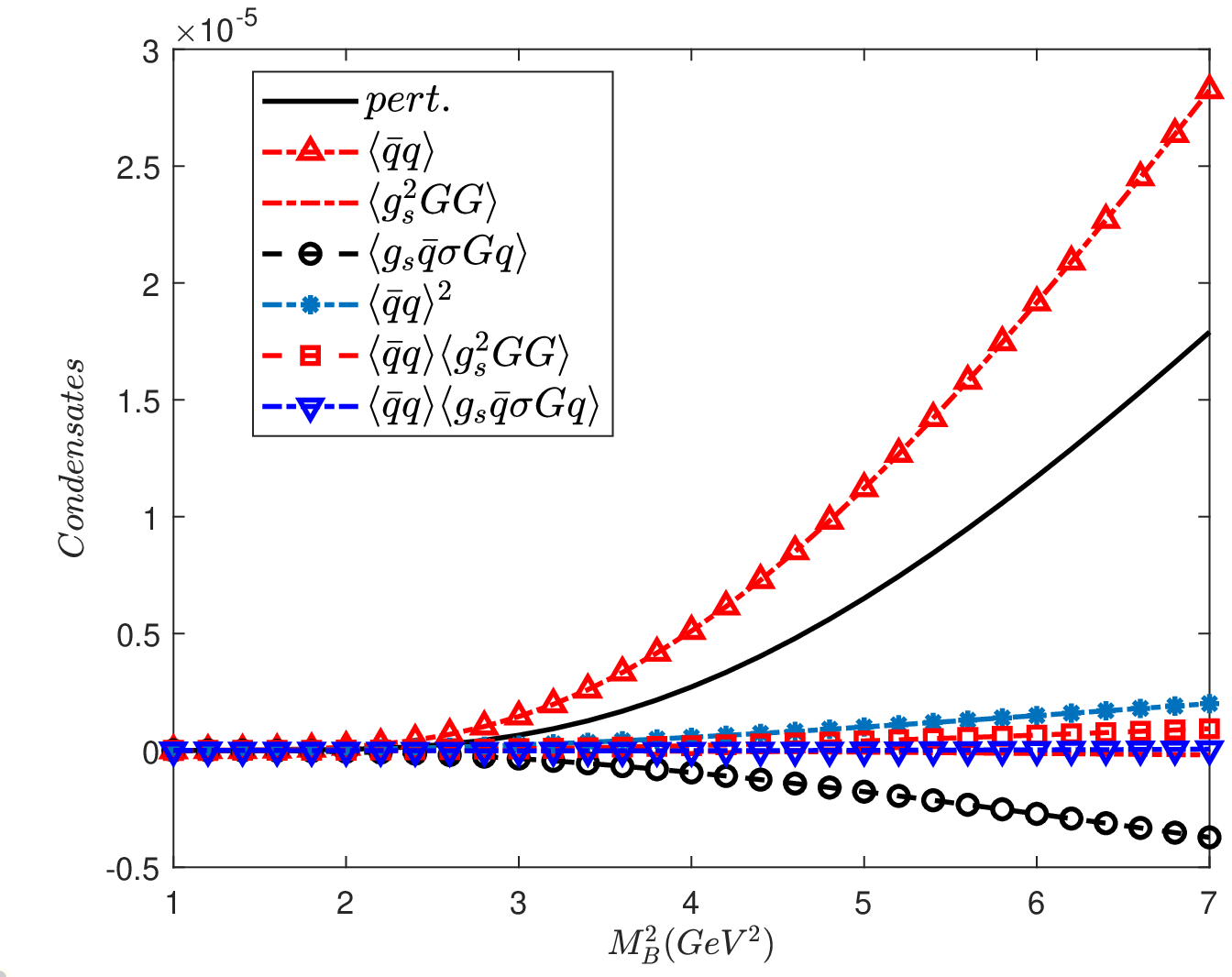}}
\subfigure[]{
\includegraphics[width=7cm]{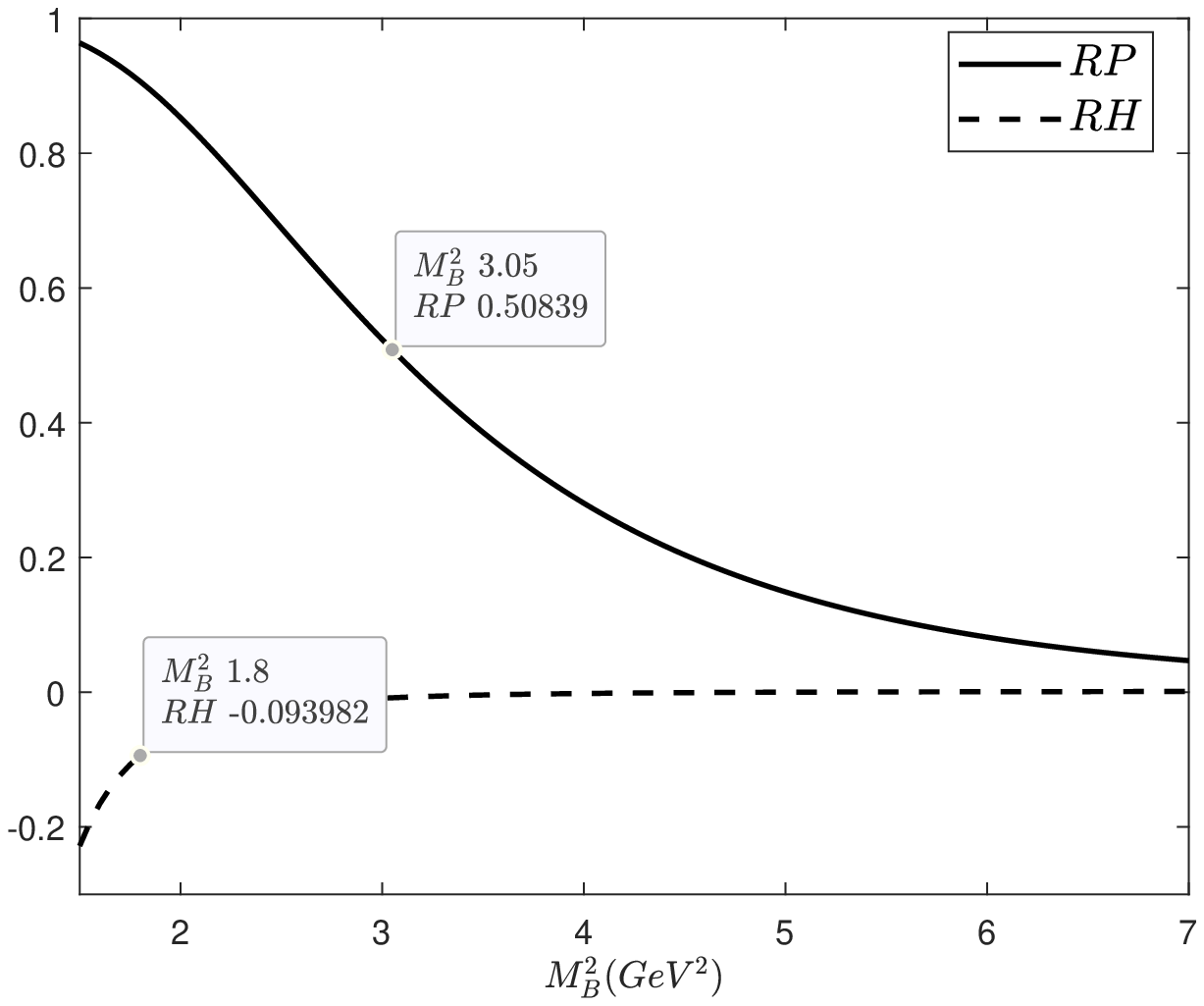}}
\caption{(a) denotes the various condensates as functions of $M^{2}_{B}$ with $\sqrt{s^{0}_{Z_{cs}}}=4.5~\mbox{GeV}$; (b) represents $RP_{0}$ and $RH_{0}$ varying with $M^{2}_{B}$ at $\sqrt{s^{0}_{Z_{cs}}}=4.5~\mbox{GeV}$.}\label{MB_range}
\end{figure}

In Fig.\ref{MB_range}(a), we compare the various terms in the OPE series as functions of $M^{2}_{B}$ with $\sqrt{s^{0}_{Z_{cs}}}=4.5~\mbox{GeV}$. From it one can see that, except the quark condensate $\langle\bar{q}q\rangle$, other vacuum condensates are much smaller than the perturbative term. So the OPE series are under control. Figure \ref{MB_range}(b) shows $RP_{0}$ and $RH_{0}$ varying with $M^{2}_{B}$ at $\sqrt{s^{0}_{Z_{cs}}}=4.5~\mbox{GeV}$. The figure shows that the requirement $RP_{0}>50\%$ gives $M^{2}_{B}\leq3.05~\mbox{GeV}^{2}$ and $|RH_{0}|<10\%$ when $M^{2}_{B}\geq 1.80~\mbox{GeV}^{2}$, which is the lower limit of $M^{2}_{B}$.

\begin{figure}[htb]
\subfigure[]{
\includegraphics[width=7cm]{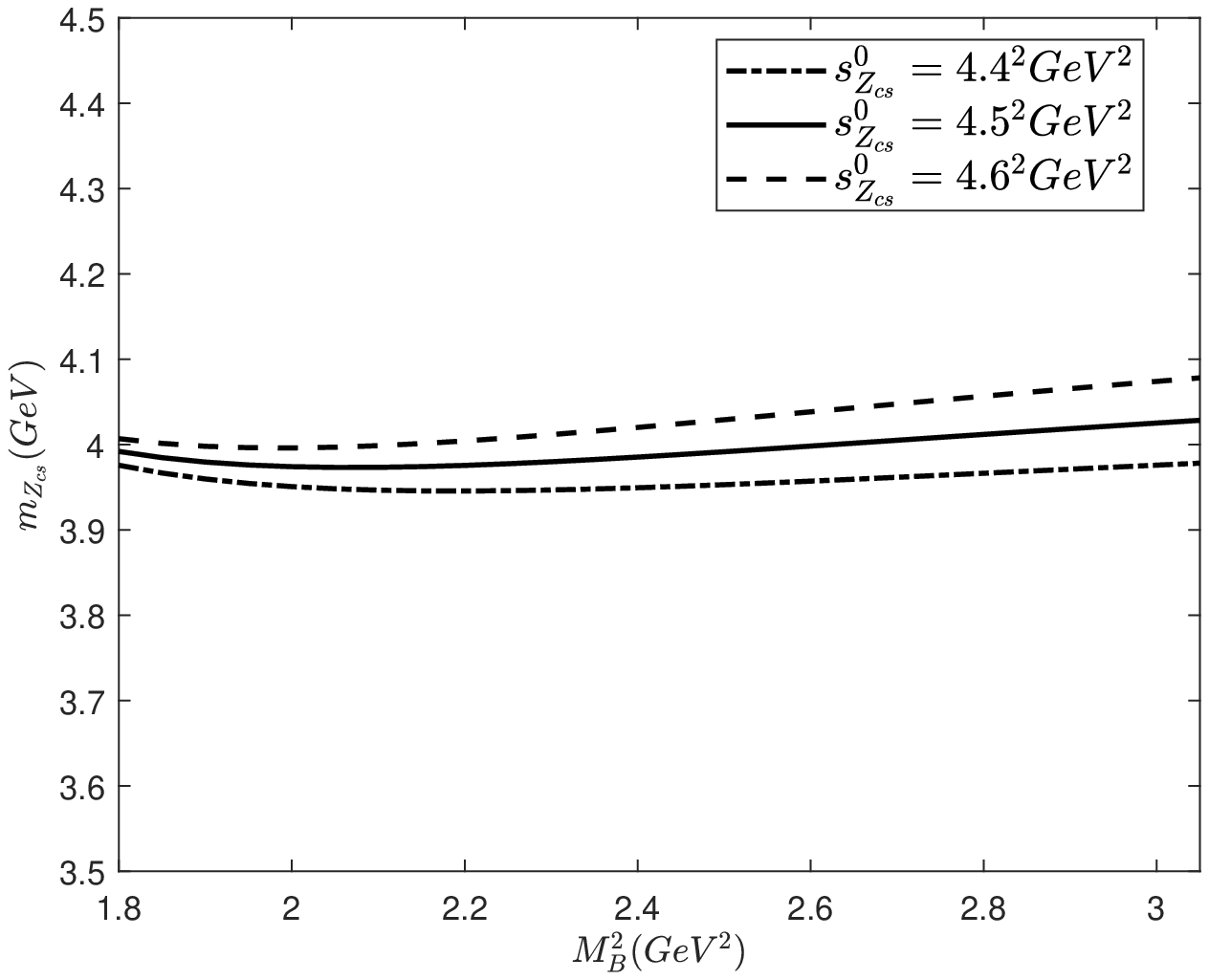}}
\subfigure[]{
\includegraphics[width=7cm]{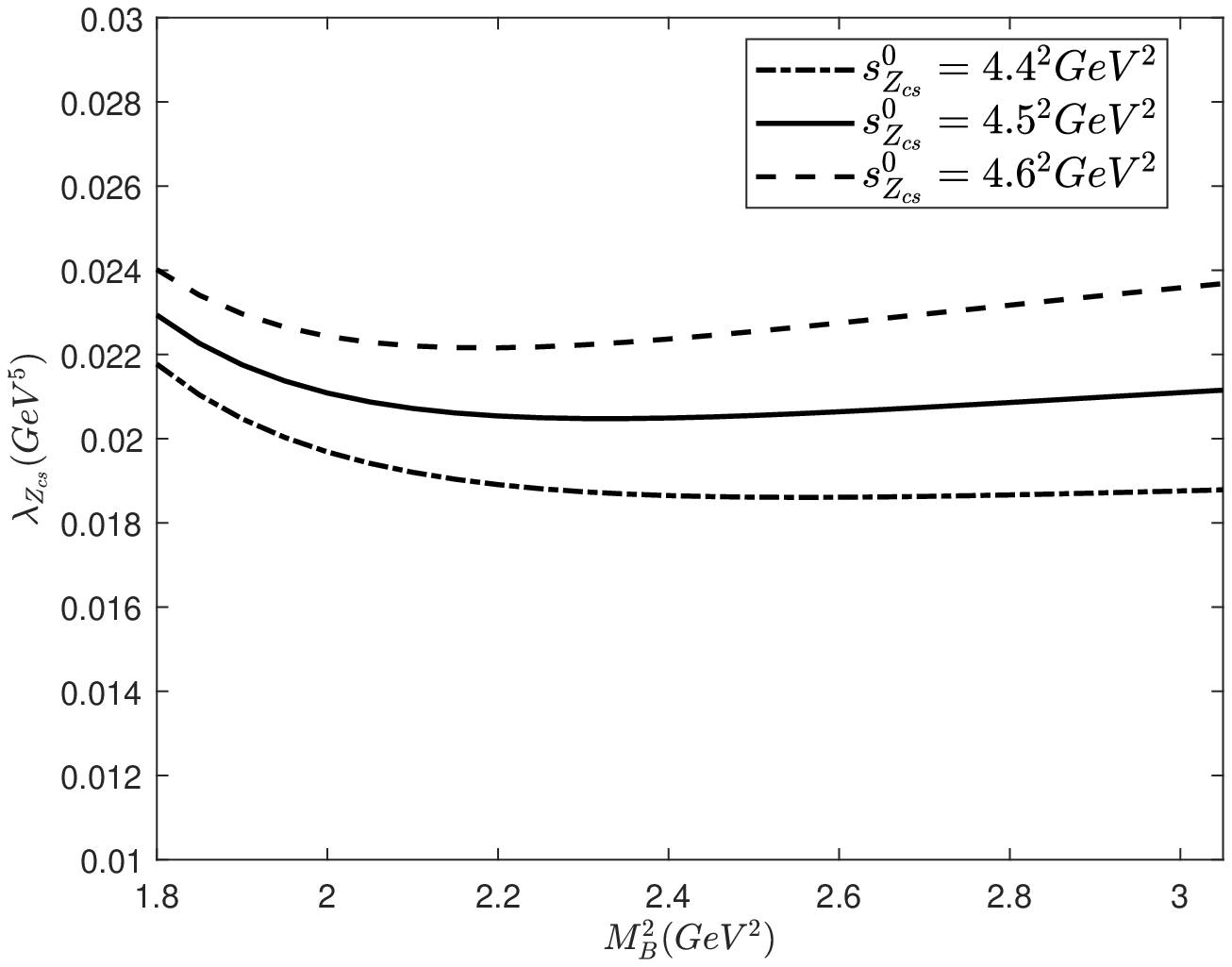}}
\caption{The dependence of the mass $m_{Z_{cs}}$ and pole residue $\lambda_{Z_{cs}}$ on the Borel parameter $M^{2}_{B}$ with $\sqrt{s^{0}_{Z_{cs}}}=4.5~\mbox{GeV}$ (dot-dashed line), $\sqrt{s^{0}_{Z_{cs}}}=4.6~\mbox{GeV}$ (real line), and $\sqrt{s^{0}_{Z_{cs}}}=4.6~\mbox{GeV}$ (dashed line).}\label{massfig}
\end{figure}

 Figure \ref{massfig} shows the dependence of $m_{Z_{cs}}$ and $\lambda_{Z_{cs}}$ on $M^{2}_{B}$ in the interval determined above with $\sqrt{s^{0}_{Z_{cs}}}=4.4~\mbox{GeV}$ (dot-dashed line), $\sqrt{s^{0}_{Z_{cs}}}=4.5~\mbox{GeV}$ (real line), and $\sqrt{s^{0}_{Z_{cs}}}=4.6~\mbox{GeV}$ (dashed line). From the figure, we can learn that the mass and pole residue vary weakly with $M^{2}_{B}$ and $s^{0}_{Z_{cs}}$. As a result, we can reliably read the value of the mass and pole residue: $m_{Z_{cs}}=3.99^{+0.17}_{-0.14}~\mbox{GeV}$ and $\lambda_{Z_{cs}}=2.07^{+0.28}_{-0.16}\times10^{-2}~\mbox{GeV}^{5}$, respectively.

The same procedure can be done for the sum rules of the magnetic moment and strong decay form factors. The value of $G_{M}(0)$ is $G_{M}(0)=0.77^{+0.66}_{-0.37}$. Finally, we obtain
\begin{equation}
\mu_{Z_{cs}}=G_{M}(0)\frac{e}{2m_{Z_{cs}}}=0.77^{+0.66}_{-0.37}\frac{e}{2m_{Z_{cs}}}=0.18^{+0.16}_{-0.09}~\mu_{N},
\end{equation}
where $\mu_{N}$ is the nucleon magneton. This result can be confronted with experimental data in the future and give important information about the inner structure of the $Z_{cs}(3985)$ state.

We summarize our results about the strong decays of $Z_{(\pm)cs}$ to $\eta_{c}K^{*}$, $J/\psi K$, $\bar{D}D^{*}_{s}$, and $\bar{D}^{*}D_{s}$ in Table \ref{results}. The processes $Z_{(-)cs}\rightarrow \eta_{c}K^{*}$ and $Z_{(-)cs}\rightarrow J/\psi K$ are suppressed in the present model as we point out in Appendix \ref{appendix}. The summed widths of the four dominant decay modes are $\Gamma_{Z_{(+)cs}}=17.47^{+12.70}_{-8.08}$ and $\Gamma_{Z_{(-)cs}}=13.86^{+10.37}_{-6.51}$ corresponding to $Z_{(+)cs}$ and $Z_{(-)cs}$, respectively, which are compatible with the experimental value.
\begin{table}[htb]
\caption{The numerical results of strong decay form factors and decay widths.}\label{results}
\begin{tabular}{|c|c|c|c|c|c|c|}
  \hline
  \hline
  Decay mode & Form factor(GeV) & $t^{0}_{1}(\mbox{GeV}^{2})$ & $t^{0}_{2}(\mbox{GeV}^{2})$ & $M^{2}_{B1}(\mbox{GeV}^{2})$ & $M^{2}_{B2}(\mbox{GeV}^{2})$ & Decay width(MeV) \\
  \hline
  $Z_{(+)cs}\rightarrow \eta_{c}K^{*}$ & $g_{1(+)}=1.76\pm0.41$ & $3.5^{2}$ & $1.4^{2}$ & [3.2,~3.8] & [0.8,~1.0] &  $3.22^{+1.68}_{-1.32}$                      \\
  \hline
  $Z_{(+)cs}\rightarrow J/\psi K$ & $g_{2(+)}=-0.47^{+0.19}_{-0.30}$ & $3.6^{2}$ & $1.0$ & [2.6,~3.0] & [0.7,~0.9] & $0.39^{+0.65}_{-0.25}$ \\
  \hline
  $Z_{(\pm)cs}\rightarrow \bar{D}D^{*}_{s}$ & $g_{3(\pm)}=4.74^{+1.30}_{-1.12}$ & $2.4^{2}$ & $2.6^{2}$ & [2.0,~2.4] & [2.1,~2.5] & $7.10^{+4.43}_{-2.96}$\\
  \hline
  $Z_{(\pm)cs}\rightarrow \bar{D}^{*}D_{s}$ & $g_{4(\pm)}=4.34^{+1.65}_{-1.33}$ & $2.5^{2}$ & $2.5^{2}$ & [2.0,~2.4] & [2.0,~2.4] & $6.76^{+5.94}_{-3.55}$\\
  \hline
  \hline
\end{tabular}
\end{table}

\section{Conclusion}\label{sec4}

In this paper, we construct $\bar{D}^{(*)}_{s}D^{(*)}$-molecule-type interpolating currents $J_{(\pm)\mu}(x)$ with $J^{P}=1^{+}$, calculate the corresponding mass and magnetic moment using the QCD sum rule method and its extension in the weak electromagnetic field, and study the processes of $Z_{(\pm)cs}$ to $\eta_{c}K^{*}$, $J/\psi K$, $\bar{D}D^{*}_{s}$, and $\bar{D}^{*}D_{s}$ via three-point sum rules. Starting with the two-point correlation function in the external electromagnetic field and expanding it in power of the electromagnetic interaction Hamiltonian, we extract the masses and pole residues of $Z_{(\pm)cs}$ states from the leading term in the expansion and the magnetic moments from the linear response to the external electromagnetic field. The numerical values are $m_{Z_{(\pm)cs}}=3.99^{+0.17}_{-0.14}~\mbox{GeV}$, $\lambda_{Z_{(\pm)cs}}=2.07^{+0.28}_{-0.16}\times10^{-2}~\mbox{GeV}^{5}$, and $\mu_{Z_{(\pm)cs}}=0.18^{+0.16}_{-0.09}~\mu_{N}$ with $\mu_{N}$ the nucleon magneton, $\Gamma_{Z_{(+)cs}}=17.47^{+12.70}_{-8.08}$, and $\Gamma_{Z_{(-)cs}}=13.86^{+10.37}_{-6.51}$. The masses are in agreement with the recently measured value of $Z_{cs}(3985)$ by the BESIII Collaboration, $m^{exp}_{Z_{cs}}=(3982.5^{+1.8}_{-2.6}\pm2.1)~\mbox{MeV}$. The widths are compatible with the experimental value $\Gamma^{exp}_{Z_{cs}}=(12.8^{+5.3}_{-4.4}\pm3.0)~\mbox{MeV}$. The pole residue is a necessary input parameter when studying other properties of the corresponding state by the QCD sum rule method or light-cone sum rule method. The magnetic moment and the various decay modes can help us to determine the inner structure of $Z_{cs}(3985)$ when being confronted with experimental data in the future.

\acknowledgments  This work was supported by the National
Natural Science Foundation of China under Contract No.~11675263.

\begin{appendix}
\section{THE QUARK PROPAGATORS}\label{appendix1}
The full quark propagators are
\begin{eqnarray}
 S^{q}_{ij}(x)=&&\frac{i \not\!{x}}{2\pi^{2}x^4}\delta_{ij}-\frac{m_{q}}{4\pi^2x^2}\delta_{ij}-\frac{\langle\bar{q}q\rangle}{12}\delta_{ij}
 +i\frac{\langle\bar{q}q\rangle}{48}m_{q}\not\!{x}\delta_{ij}-\frac{x^2}{192}\langle g_{s}\bar{q}\sigma Gq\rangle \delta_{ij}\nonumber\\
 &&+i\frac{x^2\not\!{x}}{1152}m_{q}\langle g_{s}\bar{q}\sigma Gq\rangle \delta_{ij}-i\frac{g_{s}t^{a}_{ij}G^{a}_{\mu\nu}}{32\pi^2x^2}(\not\!{x}\sigma^{\mu\nu}+\sigma^{\mu\nu}\not\!{x})\nonumber\\
 &&+i\frac{\delta_{ij}e_{q}F_{\mu\nu}}{32\pi^2x^2}(\not\!{x}\sigma^{\mu\nu}+\sigma^{\mu\nu}\not\!{x})
 -\frac{\delta_{ij}e_{q}\chi\langle\bar{q}q\rangle\sigma^{\mu\nu}F_{\mu\nu}}{24}\nonumber\\
 &&+\frac{\delta_{ij}e_{q}\langle\bar{q}q\rangle F_{\mu\nu}}{288}(\sigma^{\mu\nu}-2\sigma^{\alpha\mu}x_{\alpha}x^{\nu})\nonumber\\
 &&+\frac{\delta_{ij}e_{q}\langle\bar{q}q\rangle F_{\mu\nu}}{576}[(\kappa+\xi)\sigma^{\mu\nu}x^{2}-(2\kappa-\xi)\sigma^{\alpha\mu}x_{\alpha}x^{\nu}]+\cdots
 \end{eqnarray}
 for light quarks, and
 \begin{eqnarray}
 S^{Q}_{ij}(x)=i\int\frac{d^{4}k}{(2\pi)^4}e^{-ikx}&&[\frac{\not\!{k}+m_{Q}}{k^2-m^{2}_{Q}}\delta_{ij}
 -\frac{g_{s}t^{a}_{ij}G^{a}_{\mu\nu}}{4}\frac{\sigma^{\mu\nu}(\not\!{k}+m_{Q})+(\not\!{k}+m_{Q})\sigma^{\mu\nu}}
 {(k^2-m^{2}_{Q})^{2}}\nonumber\\
 &&+\frac{\langle g^{2}_{s}GG\rangle}{12}\delta_{ij}m_{Q}\frac{k^2+m_{Q}\not\!{k}}{(k^2-m^{2}_{Q})^{4}}\nonumber\\
 &&+\frac{\delta_{ij}e_{Q}F_{\mu\nu}}{4}\frac{\sigma^{\mu\nu}(\not\!{k}+m_{Q})+(\not\!{k}+m_{Q})\sigma^{\mu\nu}}
 {(k^2-m^{2}_{Q})^{2}}+\cdots]
 \end{eqnarray}
 for heavy quarks. In these expressions, $t^{a}=\frac{\lambda^{a}}{2}$ and $\lambda^{a}$ are the Gell-Mann matrix, $g_{s}$ is the strong interaction coupling constant, $i$ and $j$ are color indices, $e_{Q(q)}$ is the charge of the heavy (light) quark, and $F_{\mu\nu}$ is the external electromagnetic field.

\section{THE SPECTRAL DENSITIES}\label{appendix}

\subsection{The spectral densities of the two-point correlation function}
On the QCD side, we carry out the OPE up to dimension 8 for the spectral densities $\rho^{(0)}(s)$ and $\rho^{(1)}(s)$. The explicit expressions of the spectral densities are given below:

\begin{equation}
\rho^{(0)}_{(\pm)}(s)=\rho^{(d=0)}_{0}+\rho^{(d=3)}_{0}(s)+\rho^{(d=4)}_{0}(s)+\rho^{(d=5)}_{0}(s)+\rho^{(d=6)}_{0}(s)
+\rho^{(d=7)}_{0}(s)+\rho^{(d=8)}_{0}(s),
\end{equation}
with
\begin{eqnarray}
\rho^{(d=0)}_{0}(s)=&&\frac{3}{4096\pi^{6}}\int^{a_{max}}_{a_{min}}da\int^{1-a}_{b_{min}}db\frac{1}{a^{3}b^{3}}(1-a-b)(1+a+b)(m^{2}_{c}(a+b)-abs)^{4}
\nonumber\\&&-\frac{3m_{c}m_{s}}{2048\pi^{6}}\int^{a_{max}}_{a_{min}}da\int^{1-a}_{b_{min}}db\frac{1}{a^{3}b^{2}}(1-a-b)(3+a+b)(m^{2}_{c}(a+b)-abs)^{3},
\end{eqnarray}
\begin{eqnarray}
\rho^{(d=3)}_{0}(s)=&&-\frac{3m_{c}(\langle\bar{q}q\rangle+\langle\bar{s}s\rangle)}{256\pi^{4}}\int^{a_{max}}_{a_{min}}da\int^{1-a}_{b_{min}}db\frac{1}{ab^{2}}(1+a+b)(m^{2}_{c}(a+b)-abs)^{2}
\nonumber\\&&+\frac{3m^{2}_{c}m_{s}\langle\bar{q}q\rangle}{64\pi^{4}}\int^{a_{max}}_{a_{min}}da\int^{1-a}_{b_{min}}db\frac{1}{ab}(m^{2}_{c}(a+b)-abs)
\nonumber\\&&-\frac{3m_{s}\langle\bar{s}s\rangle}{256\pi^{4}}\int^{a_{max}}_{a_{min}}da\int^{1-a}_{b_{min}}db\frac{1}{ab}(m^{2}_{c}(a+b)-abs)^{2}
\nonumber\\&&+\frac{3m_{s}\langle\bar{s}s\rangle}{256\pi^{4}}\int^{a_{max}}_{a_{min}}da\frac{1}{a(1-a)}(m^{2}_{c}-a(1-a)s)^{2},
\end{eqnarray}
\begin{eqnarray}
\rho^{(d=4)}_{0}(s)=&&\frac{m^{2}_{c}\langle g^{2}_{s}GG\rangle}{2048\pi^{6}}\int^{a_{max}}_{a_{min}}da\int^{1-a}_{b_{min}}db\frac{1}{a^{3}}(1-a-b)(1+a+b)(m^{2}_{c}(a+b)-abs)
\nonumber\\&&-\frac{\langle g^{2}_{s}GG\rangle}{2048\pi^{6}}\int^{a_{max}}_{a_{min}}da\int^{1-a}_{b_{min}}db\frac{1}{a^{2}b}(1-2a-2b)(m^{2}_{c}(a+b)-abs)^{2}
\nonumber\\&&-\frac{3m_{c}m_{s}\langle g^{2}_{s}GG\rangle}{8192\pi^{6}}\int^{a_{max}}_{a_{min}}da\int^{1-a}_{b_{min}}db\frac{1}{a^{3}}(1-a-b)(3+a+b)(m^{2}_{c}(a+b)-abs)
\nonumber\\&&-\frac{m^{3}_{c}m_{s}\langle g^{2}_{s}GG\rangle}{8192\pi^{6}}\int^{a_{max}}_{a_{min}}da\int^{1-a}_{b_{min}}db\frac{1}{a^{3}}(1-a-b)(3+a+b)(a+b)
\nonumber\\&&-\frac{m_{c}m_{s}\langle g^{2}_{s}GG\rangle}{4096\pi^{6}}\int^{a_{max}}_{a_{min}}da\int^{1-a}_{b_{min}}db\frac{1}{ab}(1+a+b)(m^{2}_{c}(a+b)-abs),
\end{eqnarray}
\begin{eqnarray}
\rho^{(d=5)}_{0}(s)=&&\frac{3m_{c}(\langle g_{s}\bar{q}\sigma Gq\rangle+\langle g_{s}\bar{s}\sigma Gs\rangle)}{256\pi^{4}}\int^{a_{max}}_{a_{min}}da\int^{1-a}_{b_{min}}db\frac{1}{a^{2}}(a+b)(m^{2}_{c}(a+b)-abs)\nonumber\\
&&+\frac{3m_{c}(\langle g_{s}\bar{q}\sigma Gq\rangle+\langle g_{s}\bar{s}\sigma Gs\rangle)}{512\pi^{4}}\int^{a_{max}}_{a_{min}}da\int^{1-a}_{b_{min}}db\frac{1}{b}(m^{2}_{c}(a+b)-abs)\nonumber\\
&&-\frac{3m^{2}_{c}m_{s}\langle g_{s}\bar{q}\sigma Gq\rangle}{256\pi^{4}}\int^{a_{max}}_{a_{min}}da\int^{1-a}_{b_{min}}db\frac{1}{a}\nonumber\\&&-\frac{3m_{c}(\langle g_{s}\bar{q}\sigma Gq\rangle+\langle g_{s}\bar{s}\sigma Gs\rangle)}{256\pi^{4}}\int^{a_{max}}_{a_{min}}da\frac{1}{a}(m^{2}_{c}-a(1-a)s)\nonumber\\&&
+\frac{3m^{2}_{c}m_{s}\langle g_{s}\bar{q}\sigma Gq\rangle}{256\pi^{4}}\int^{a_{max}}_{a_{min}}da\nonumber\\&&
+\frac{m_{s}\langle g_{s}\bar{s}\sigma Gs\rangle}{256\pi^{4}}\int^{a_{max}}_{a_{min}}da(m^{2}_{c}-2a(1-a)s),
\end{eqnarray}
\begin{eqnarray}
\rho^{(d=6)}_{0}(s)=&&\frac{m^{2}_{c}\langle\bar{q}q\rangle\langle\bar{s}s\rangle}{16\pi^{2}}\frac{\sqrt{s(s-4m^{2}_{c})}}{s}
-\frac{3m_{c}m_{s}\langle\bar{q}q\rangle\langle\bar{s}s\rangle}{128\pi^{2}}\frac{\sqrt{s(s-4m^{2}_{c})}}{s}\nonumber\\&&
-\frac{m^{3}_{c}m_{s}\langle\bar{q}q\rangle\langle\bar{s}s\rangle}{32\pi^{2}}\frac{1}{\sqrt{s(s-4m^{2}_{c})}},
\end{eqnarray}
\begin{eqnarray}
\rho^{(d=7)}_{0}(s)=&&-\frac{m_{c}(\langle\bar{q}q\rangle+\langle\bar{s}s\rangle)\langle g^{2}_{s}GG\rangle}{1024\pi^{4}}\int^{a_{max}}_{a_{min}}da\int^{1-a}_{b_{min}}db\frac{1}{a^2}b(1+a+b)\nonumber\\&&
+\frac{m_{c}(\langle\bar{q}q\rangle+\langle\bar{s}s\rangle)\langle g^{2}_{s}GG\rangle}{3072\pi^{4}}\int^{a_{max}}_{a_{min}}da\int^{1-a}_{b_{min}}db\nonumber\\&&
-\frac{m_{c}(\langle\bar{q}q\rangle+\langle\bar{s}s\rangle)\langle g^{2}_{s}GG\rangle}{1536\pi^{4}}\int^{a_{max}}_{a_{min}}da\nonumber\\
&&+\frac{m^{3}_{c}(\langle\bar{q}q\rangle+\langle\bar{s}s\rangle)\langle g^{2}_{s}GG\rangle}{3072\pi^{4}}\int^{a_{max}}_{a_{min}}da\frac{(1+a+b_{min})b_{min}}{a(as-m^{2}_{c})}\nonumber\\&&
+\frac{m^{3}_{c}(\langle\bar{q}q\rangle+\langle\bar{s}s\rangle)\langle g^{2}_{s}GG\rangle}{3072\pi^{4}}\int^{a_{max}}_{a_{min}}da\frac{(1+a+b_{min})b^{2}_{min}}{a^{2}(as-m^{2}_{c})}\nonumber\\&&
+\frac{m^{2}_{c}m_{s}\langle\bar{s}s\rangle\langle g^{2}_{s}GG\rangle}{1536\pi^{4}}\int^{a_{max}}_{a_{min}}da\frac{b^{2}_{min}}{a(as-m^{2}_{c})}\nonumber\\&&
-\frac{m^{2}_{c}m_{s}\langle\bar{q}q\rangle\langle g^{2}_{s}GG\rangle}{256\pi^{4}}\int^{a_{max}}_{a_{min}}da\frac{b_{min}}{a(as-m^{2}_{c})}\nonumber\\&&
+\frac{m^{4}_{c}m_{s}\langle\bar{q}q\rangle\langle g^{2}_{s}GG\rangle}{768\pi^{4}M^{2}_{B}}\int^{a_{max}}_{a_{min}}da\frac{b_{min}}{a^{2}(as-m^{2}_{c})}\nonumber\\&&
-\frac{m^{2}_{c}m_{s}\langle\bar{s}s\rangle\langle g^{2}_{s}GG\rangle}{1536\pi^{4}}\frac{1}{\sqrt{s(s-4m^{2}_{c})}}(\frac{a^{2}_{min}}{a_{max}}+\frac{a^{2}_{max}}{a_{min}})
\nonumber\\&&+\frac{m^{2}_{c}m_{s}\langle\bar{s}s\rangle\langle g^{2}_{s}GG\rangle}{3072\pi^{4}}\frac{1}{\sqrt{s(s-4m^{2}_{c})}},
\end{eqnarray}
\begin{eqnarray}
\rho^{(d=8)}_{0}(s)=&&-\frac{(\langle\bar{q}q\rangle\langle g_{s}\bar{s}\sigma Gs\rangle+\langle\bar{s}s\rangle\langle g_{s}\bar{q}\sigma Gq\rangle)}{64\pi^2}\times\nonumber\\&&\frac{m^{2}_{c}}{\sqrt{s(s-4m^{2}_{c})}}(\frac{2m^{2}_{c}}{M^{2}_{B}}+\frac{2m^{2}_{c}}{s}-1)
\nonumber\\&&+\frac{(2\langle\bar{q}q\rangle\langle g_{s}\bar{s}\sigma Gs\rangle+3\langle\bar{s}s\rangle\langle g_{s}\bar{q}\sigma Gq\rangle)}{768\pi^2}\times\nonumber\\&&\frac{m^{3}_{c}m_{s}}{\sqrt{s(s-4m^{2}_{c})}}(\frac{2s}{M^{4}_{B}}+\frac{3}{M^{2}_{B}}+\frac{2}{s})
\nonumber\\&&-\frac{\langle\bar{s}s\rangle\langle g_{s}\bar{q}\sigma Gq\rangle}{128\pi^2M^{2}_{B}}\frac{m_{c}m_{s}s}{\sqrt{s(s-4m^{2}_{c})}}(a^{2}_{max}+a^{2}_{min}),
\end{eqnarray}

\begin{equation}
\rho^{(1)}_{(\pm)}(s)=\rho^{(d=2)}_{1}+\rho^{(d=3)}_{1}(s)+\rho^{(d=5)}_{1}(s)+\rho^{(d=6)}_{1}(s)
+\rho^{(d=7)}_{1}(s)+\rho^{(d=8)}_{1}(s),
\end{equation}
with
\begin{eqnarray}
\rho^{(d=2)}_{1}(s)=&&\frac{3(e_{s}-e_{u})}{1024\pi^{6}}\int^{a_{max}}_{a_{min}}da\int^{1-a}_{b_{min}}db\frac{1}{a^{2}b^{2}}(a+b)(m^{2}_{c}(a+b)-abs)^{3}
\nonumber\\&&+\frac{9e_{u}m_{c}m_{s}}{1024\pi^{6}}\int^{a_{max}}_{a_{min}}da\int^{1-a}_{b_{min}}db\frac{1}{ab^{2}}(a+b)(m^{2}_{c}(a+b)-abs)^{2}
\nonumber\\&&-\frac{9e_{c}m_{c}m_{s}}{2048\pi^{6}}\int^{a_{max}}_{a_{min}}da\int^{1-a}_{b_{min}}db\frac{1}{a^{2}b^{2}}(1-a-b)(1+a+b)(m^{2}_{c}(a+b)-abs)^{2}
\nonumber\\&&+\frac{9e_{c}m_{c}m_{s}}{1024\pi^{6}}\int^{a_{max}}_{a_{min}}da\int^{1-a}_{b_{min}}db\frac{1}{a^{3}b}(1-a-b)(m^{2}_{c}(a+b)-abs)^{2},
\end{eqnarray}
\begin{eqnarray}
\rho^{(d=3)}_{1}(s)=&&\frac{3m_{c}\chi(e_{u}\langle\bar{q}q\rangle-e_{s}\langle\bar{s}s\rangle)}{256\pi^{4}}\int^{a_{max}}_{a_{min}}da\int^{1-a}_{b_{min}}db\frac{1}{a^{2}b}(m^{2}_{c}(a+b)-abs)^{2}
\nonumber\\&&-\frac{3m_{s}\chi e_{s}\langle\bar{s}s\rangle}{256\pi^{4}}\int^{a_{max}}_{a_{min}}da\int^{1-a}_{b_{min}}db\frac{1}{ab}(m^{2}_{c}(a+b)-abs)^{2}
\nonumber\\&&-\frac{3m^{2}_{c}m_{s}\chi e_{u}\langle\bar{q}q\rangle}{128\pi^{4}}\int^{a_{max}}_{a_{min}}da\int^{1-a}_{b_{min}}db\frac{1}{ab}(m^{2}_{c}(a+b)-abs)
\nonumber\\&&+\frac{3m_{s}\chi e_{s}\langle\bar{s}s\rangle}{256\pi^{4}}\int^{a_{max}}_{a_{min}}da\frac{1}{a(1-a)}(m^{2}_{c}-a(1-a)s)^{2},
\end{eqnarray}
\begin{eqnarray}
\rho^{(d=5)}_{1}(s)=&&\frac{3m_{c}e_{c}(\langle\bar{q}q\rangle-\langle\bar{s}s\rangle)}{128\pi^{4}}\int^{a_{max}}_{a_{min}}da\int^{1-a}_{b_{min}}db\frac{1}{ab}(a+b)(m^{2}_{c}(a+b)-abs)
\nonumber\\&&-\frac{3m_{c}e_{c}(\langle\bar{q}q\rangle-\langle\bar{s}s\rangle)}{128\pi^{4}}\int^{a_{max}}_{a_{min}}da\int^{1-a}_{b_{min}}db\frac{1}{a^{2}}(m^{2}_{c}(a+b)-abs)
\nonumber\\&&+\frac{3m_{c}(e_{s}\langle\bar{q}q\rangle-e_{u}\langle\bar{s}s\rangle)}{128\pi^{4}}\int^{a_{max}}_{a_{min}}da\int^{1-a}_{b_{min}}db\frac{1}{b}(m^{2}_{c}(a+b)-abs)
\nonumber\\&&-\frac{m_{c}(e_{u}\langle\bar{q}q\rangle-e_{s}\langle\bar{s}s\rangle)}{128\pi^{4}}\int^{a_{max}}_{a_{min}}da\int^{1-a}_{b_{min}}db\frac{1}{b}(m^{2}_{c}(a+b)-abs)
\nonumber\\&&+\frac{m^{2}_{c}m_{s}e_{u}\langle\bar{q}q\rangle}{128\pi^{4}}\int^{a_{max}}_{a_{min}}da\int^{1-a}_{b_{min}}db
\nonumber\\&&-\frac{m_{c}(2\kappa-\xi)(e_{u}\langle\bar{q}q\rangle-e_{s}\langle\bar{s}s\rangle)}{512\pi^{4}}\int^{a_{max}}_{a_{min}}da\int^{1-a}_{b_{min}}db\frac{1}{b}(m^{2}_{c}(a+b)-abs)\nonumber\\
&&+\frac{m^{2}_{c}m_{s}(2\kappa-\xi)e_{u}\langle\bar{q}q\rangle}{512\pi^{4}}\int^{a_{max}}_{a_{min}}da\int^{1-a}_{b_{min}}db\nonumber\\
&&+\frac{3m_{c}(2\kappa+\xi)(e_{u}\langle\bar{q}q\rangle-e_{s}\langle\bar{s}s\rangle)}{256\pi^{4}}\int^{a_{max}}_{a_{min}}da\int^{1-a}_{b_{min}}db\frac{1}{b^{2}}(m^{2}_{c}(a+b)-abs)\nonumber\\
&&-\frac{3m^{2}_{c}m_{s}(2\kappa+\xi)e_{u}\langle\bar{q}q\rangle}{256\pi^{4}}\int^{a_{max}}_{a_{min}}da\int^{1-a}_{b_{min}}db\frac{1}{b}\nonumber\\
&&-\frac{3m_{c}(e_{s}\langle\bar{q}q\rangle-e_{u}\langle\bar{s}s\rangle)}{128\pi^{4}}\int^{a_{max}}_{a_{min}}da\frac{1}{a}(m^{2}_{c}-a(1-a)s)\nonumber\\
&&-\frac{3m_{s}e_{u}\langle\bar{s}s\rangle}{256\pi^{4}}\int^{a_{max}}_{a_{min}}da(m^{2}_{c}-2a(1-a)s)\nonumber\\
&&-\frac{m_{c}(e_{u}\langle\bar{q}q\rangle-e_{s}\langle\bar{s}s\rangle)}{128\pi^{4}}\int^{a_{max}}_{a_{min}}da\frac{1}{a}(m^{2}_{c}-a(1-a)s)\nonumber\\
&&+\frac{m^{2}_{c}m_{s}e_{u}\langle\bar{q}q\rangle}{128\pi^{4}}\int^{a_{max}}_{a_{min}}da\nonumber\\
&&-\frac{m_{c}(\kappa+\xi)(e_{u}\langle\bar{q}q\rangle-e_{s}\langle\bar{s}s\rangle)}{256\pi^{4}}\int^{a_{max}}_{a_{min}}da\frac{1}{a}(m^{2}_{c}-a(1-a)s)\nonumber\\
&&+\frac{m^{2}_{c}m_{s}(\kappa+\xi)e_{u}\langle\bar{q}q\rangle}{256\pi^{4}}\int^{a_{max}}_{a_{min}}da,
\end{eqnarray}
\begin{eqnarray}
\rho^{(d=6)}_{1}(s)=&&-\frac{3m_{c}m_{s}e_{c}\langle g^{2}_{s}GG\rangle}{8192\pi^{6}}\int^{a_{max}}_{a_{min}}da\int^{1-a}_{b_{min}}db\frac{1}{a^{2}}(1-a-b)(1+a+b)\nonumber\\
&&+\frac{3m_{c}m_{s}e_{c}\langle g^{2}_{s}GG\rangle}{4096\pi^{6}}\int^{a_{max}}_{a_{min}}da\int^{1-a}_{b_{min}}db\frac{1}{b}\nonumber\\
&&+\frac{m^{2}_{c}(e_{s}-e_{u})\langle g^{2}_{s}GG\rangle}{2048\pi^{6}}\int^{a_{max}}_{a_{min}}da\int^{1-a}_{b_{min}}db\frac{1}{a^{2}}b(a+b)\nonumber\\
&&+\frac{3m_{c}m_{s}e_{u}\langle g^{2}_{s}GG\rangle}{4096\pi^{6}}\int^{a_{max}}_{a_{min}}da\int^{1-a}_{b_{min}}db\frac{1}{a^{2}}b(a+b)\nonumber\\
&&-\frac{3(e_{s}-e_{u})\langle g^{2}_{s}GG\rangle}{4096\pi^{6}}\int^{a_{max}}_{a_{min}}da\int^{1-a}_{b_{min}}db\frac{1}{a}(m^{2}_{c}(a+b)-abs)\nonumber\\
&&-\frac{m^{3}_{c}m_{s}e_{c}\langle g^{2}_{s}GG\rangle}{4096\pi^{6}}\int^{a_{max}}_{a_{min}}da\frac{1-a-b_{min}}{a(as-m^{2}_{c})}\nonumber\\
&&+\frac{m^{3}_{c}m_{s}e_{c}\langle g^{2}_{s}GG\rangle}{8192\pi^{6}}\int^{a_{max}}_{a_{min}}da\frac{(1-a-b_{min})(1+a+b_{min})b_{min}}{a^{2}(as-m^{2}_{c})}\nonumber\\
&&-\frac{m^{3}_{c}m_{s}e_{u}\langle g^{2}_{s}GG\rangle}{4096\pi^{6}}\int^{a_{max}}_{a_{min}}da\frac{(a+b_{min})^{2}b_{min}}{a^{2}(as-m^{2}_{c})}\nonumber\\
&&+\frac{3(e_{s}-e_{u})\langle g^{2}_{s}GG\rangle}{4096\pi^{6}}\int^{a_{max}}_{a_{min}}da\frac{1}{a}(m^{2}_{c}-a(1-a)s)\nonumber\\
&&+\frac{m^{2}_{c}(e_{s}-e_{u})\chi\langle\bar{q}q\rangle\langle\bar{s}s\rangle}{32\pi^{2}}\int^{a_{max}}_{a_{min}}da\nonumber\\
&&-\frac{m_{c}m_{s}(e_{s}-e_{u})\chi\langle\bar{q}q\rangle\langle\bar{s}s\rangle}{32\pi^{2}}\int^{a_{max}}_{a_{min}}daa\nonumber\\
&&-\frac{m^{3}_{c}m_{s}(2e_{s}-e_{u})\chi\langle\bar{q}q\rangle\langle\bar{s}s\rangle}{64\pi^{2}}\frac{1}{\sqrt{s(s-4m^{2}_{c})}},
\end{eqnarray}
\begin{eqnarray}
\rho^{(d=7)}_{1}(s)=&&-\frac{3m_{c}e_{c}(\langle g_{s}\bar{q}\sigma Gq\rangle-\langle g_{s}\bar{s}\sigma Gs\rangle)}{512\pi^{4}}\int^{a_{max}}_{a_{min}}da\int^{1-a}_{b_{min}}db\nonumber\\
&&-\frac{3m_{c}e_{c}(\langle g_{s}\bar{q}\sigma Gq\rangle-\langle g_{s}\bar{s}\sigma Gs\rangle)}{256\pi^{4}}\int^{a_{max}}_{a_{min}}da\int^{1-a}_{b_{min}}db\frac{1}{b}(a+b)\nonumber\\
&&-\frac{3m_{c}(e_{s}\langle g_{s}\bar{q}\sigma Gq\rangle-e_{u}\langle g_{s}\bar{s}\sigma Gs\rangle)}{256\pi^{4}}\int^{a_{max}}_{a_{min}}da\int^{1-a}_{b_{min}}db\frac{a}{b}\nonumber\\
&&+\frac{m_{c}\chi(e_{u}\langle\bar{q}q\rangle-e_{s}\langle\bar{s}s\rangle)\langle g^{2}_{s}GG\rangle}{1024\pi^{4}}\int^{a_{max}}_{a_{min}}da\int^{1-a}_{b_{min}}db\frac{b}{a^{2}}\nonumber\\
&&+\frac{3m_{c}e_{c}(\langle g_{s}\bar{q}\sigma Gq\rangle-\langle g_{s}\bar{s}\sigma Gs\rangle)}{512\pi^{4}}\int^{a_{max}}_{a_{min}}da\frac{(2a-1)}{a}\nonumber\\
&&-\frac{3m^{2}_{c}m_{s}e_{c}\langle g_{s}\bar{q}\sigma Gq\rangle}{256\pi^{4}}\int^{a_{max}}_{a_{min}}da\frac{1}{as-m^{2}_{c}}\nonumber\\
&&-\frac{3m_{c}(e_{s}\langle g_{s}\bar{q}\sigma Gq\rangle-e_{u}\langle g_{s}\bar{s}\sigma Gs\rangle)}{512\pi^{4}}\int^{a_{max}}_{a_{min}}daa\nonumber\\
&&+\frac{3m_{s}e_{u}\langle g_{s}\bar{s}\sigma Gs\rangle}{512\pi^{4}}\int^{a_{max}}_{a_{min}}daa(a-1)\nonumber\\
&&-\frac{3m_{c}(e_{s}\langle g_{s}\bar{q}\sigma Gq\rangle-e_{u}\langle g_{s}\bar{s}\sigma Gs\rangle)}{256\pi^{4}}\int^{a_{max}}_{a_{min}}da\frac{a}{a-1}\nonumber\\
&&-\frac{m^{3}_{c}\chi(e_{u}\langle\bar{q}q\rangle-e_{s}\langle\bar{s}s\rangle)\langle g^{2}_{s}GG\rangle}{3072\pi^{4}}\int^{a_{max}}_{a_{min}}da\frac{b_{min}}{a(as-m^{2}_{c})}\nonumber\\
&&-\frac{m^{3}_{c}\chi(e_{u}\langle\bar{q}q\rangle-e_{s}\langle\bar{s}s\rangle)\langle g^{2}_{s}GG\rangle}{3072\pi^{4}}\int^{a_{max}}_{a_{min}}da\frac{b^{2}_{min}}{a^{2}(as-m^{2}_{c})}\nonumber\\
&&+\frac{m^{2}_{c}m_{s}\chi e_{u}\langle\bar{q}q\rangle\langle g^{2}_{s}GG\rangle}{512\pi^{4}}\int^{a_{max}}_{a_{min}}da\frac{b_{min}}{a(as-m^{2}_{c})}\nonumber\\
&&-\frac{m^{4}_{c}m_{s}\chi e_{u}\langle\bar{q}q\rangle\langle g^{2}_{s}GG\rangle}{1536\pi^{4}M^{2}_{B}}\int^{a_{max}}_{a_{min}}da\frac{b_{min}}{a^{2}(as-m^{2}_{c})}\nonumber\\
&&+\frac{m^{2}_{c}m_{s}\chi e_{s}\langle\bar{s}s\rangle\langle g^{2}_{s}GG\rangle}{1536\pi^{4}}\int^{a_{max}}_{a_{min}}da\frac{b^{2}_{min}}{a(as-m^{2}_{c})}\nonumber\\
&&+\frac{m_{c}\chi(e_{u}\langle\bar{q}q\rangle-e_{s}\langle\bar{s}s\rangle)\langle g^{2}_{s}GG\rangle}{1024\pi^{4}}\int^{a_{max}}_{a_{min}}da\nonumber\\
&&-\frac{m_{s}\chi e_{s}\langle\bar{s}s\rangle\langle g^{2}_{s}GG\rangle}{1024\pi^{4}}\int^{a_{max}}_{a_{min}}da(a-1)\nonumber\\
&&-\frac{3m^{3}_{c}(e_{s}\langle g_{s}\bar{q}\sigma Gq\rangle-e_{u}\langle g_{s}\bar{s}\sigma Gs\rangle)}{512\pi^{4}}\frac{1}{\sqrt{s(s-4m^{2}_{c})}}\nonumber\\
&&-\frac{m^{4}_{c}m_{s}e_{u}\langle g_{s}\bar{s}\sigma Gs\rangle}{256\pi^{4}}\frac{1}{\sqrt{s(s-4m^{2}_{c})}}(\frac{1}{M^{2}_{B}}+\frac{3}{s})\nonumber\\
&&-\frac{m^{2}_{c}m_{s}\chi e_{s}\langle\bar{s}s\rangle\langle g^{2}_{s}GG\rangle}{1536\pi^{4}}\frac{1}{\sqrt{s(s-4m^{2}_{c})}}(\frac{a^{2}_{max}}{a_{min}}+\frac{a^{2}_{min}}{a_{max}})\nonumber\\
&&+\frac{m^{2}_{c}m_{s}\chi e_{s}\langle\bar{s}s\rangle\langle g^{2}_{s}GG\rangle}{1024\pi^{4}}\frac{1}{\sqrt{s(s-4m^{2}_{c})}},
\end{eqnarray}
\begin{eqnarray}
\rho^{(d=8)}_{1}(s)=&&-\frac{m^{3}_{c}m_{s}e_{c}\langle\bar{q}q\rangle\langle\bar{s}s\rangle}{32\pi^{2}M^{2}_{B}}\frac{1}{\sqrt{s(s-4m^{2}_{c})}}\nonumber\\
&&+\frac{m_{c}m_{s}e_{c}\langle\bar{q}q\rangle\langle\bar{s}s\rangle}{64\pi^{2}}\frac{a^{2}_{max}+a^{2}_{min}}{\sqrt{s(s-4m^{2}_{c})}}(\frac{s}{M^{2}_{B}}+1)\nonumber\\
&&+\frac{m^{2}_{c}\chi(e_{u}\langle\bar{q}q\rangle\langle g_{s}\bar{s}\sigma Gs\rangle-e_{s}\langle\bar{s}s\rangle\langle g_{s}\bar{q}\sigma Gq\rangle)}{64\pi^{2}}\times\nonumber\\&&\frac{1}{\sqrt{s(s-4m^{2}_{c})}}(\frac{m^{2}_{c}}{M^{2}_{B}}+\frac{m^{2}_{c}}{s}-1)\nonumber\\
&&-\frac{m^{3}_{c}m_{s}\chi(e_{u}\langle\bar{q}q\rangle\langle g_{s}\bar{s}\sigma Gs\rangle-3e_{s}\langle\bar{s}s\rangle\langle g_{s}\bar{q}\sigma Gq\rangle)}{384\pi^{2}}\times\nonumber\\&&\frac{1}{\sqrt{s(s-4m^{2}_{c})}}(\frac{s}{M^{4}_{B}}+\frac{2}{M^{2}_{B}}+\frac{2}{s})\nonumber\\
&&-\frac{m^{3}_{c}m_{s}\chi e_{s}\langle\bar{s}s\rangle\langle g_{s}\bar{q}\sigma Gq\rangle}{128\pi^{2}}\frac{1}{\sqrt{s(s-4m^{2}_{c})}}(\frac{1}{M^{2}_{B}}+\frac{2}{s})\nonumber\\
&&-\frac{m_{c}m_{s}\chi e_{s}\langle\bar{s}s\rangle\langle g_{s}\bar{q}\sigma Gq\rangle}{64\pi^{2}M^{2}_{B}}\frac{s}{\sqrt{s(s-4m^{2}_{c})}}(a^{2}_{max}+a^{2}_{min})\nonumber\\
&&+\frac{m^{4}_{c}(e_{s}-e_{u})\langle\bar{q}q\rangle\langle\bar{s}s\rangle}{48\pi^{2}}\frac{1}{\sqrt{s(s-4m^{2}_{c})}}(\frac{1}{M^{2}_{B}}+\frac{2}{s})\nonumber\\
&&+\frac{m^{3}_{c}m_{s}e_{u}\langle\bar{q}q\rangle\langle\bar{s}s\rangle}{192\pi^{2}}\frac{1}{\sqrt{s(s-4m^{2}_{c})}}(\frac{s}{M^{4}_{B}}+\frac{3}{M^{2}_{B}}+\frac{4}{s})\nonumber\\
&&+\frac{m^{4}_{c}(e_{s}-e_{u})(\kappa+\xi)\langle\bar{q}q\rangle\langle\bar{s}s\rangle}{96\pi^{2}}\frac{1}{\sqrt{s(s-4m^{2}_{c})}}(\frac{1}{M^{2}_{B}}+\frac{1}{s})\nonumber\\
&&+\frac{m^{4}_{c}(e_{s}-e_{u})(2\kappa-\xi)\langle\bar{q}q\rangle\langle\bar{s}s\rangle}{192\pi^{2}}\frac{1}{s\sqrt{s(s-4m^{2}_{c})}}\nonumber\\
&&+\frac{m^{3}_{c}m_{s}e_{u}(\kappa+\xi)\langle\bar{q}q\rangle\langle\bar{s}s\rangle}{384\pi^{2}}\frac{1}{\sqrt{s(s-4m^{2}_{c})}}(\frac{s}{M^{4}_{B}}+\frac{2}{M^{2}_{B}}+\frac{2}{s})\nonumber\\
&&+\frac{m^{3}_{c}m_{s}e_{u}(2\kappa-\xi)\langle\bar{q}q\rangle\langle\bar{s}s\rangle}{768\pi^{2}}\frac{1}{\sqrt{s(s-4m^{2}_{c})}}(\frac{1}{M^{2}_{M}}+\frac{2}{s})\nonumber\\
&&-\frac{m^{2}_{c}(e_{s}-e_{u})(2\kappa+\xi)\langle\bar{q}q\rangle\langle\bar{s}s\rangle}{64\pi^{2}}\frac{1}{\sqrt{s(s-4m^{2}_{c})}}\nonumber\\
&&-\frac{m_{c}m_{s}e_{u}(2\kappa+\xi)\langle\bar{q}q\rangle\langle\bar{s}s\rangle}{128\pi^{2}}\frac{a^{2}_{max}+a^{2}_{min}}{\sqrt{s(s-4m^{2}_{c})}}(\frac{s}{M^{2}_{B}}+1).
\end{eqnarray}
In the above equations, $a_{max}=\frac{1+\sqrt{1-\frac{4m^{2}_{c}}{s}}}{2}$, $a_{min}=\frac{1-\sqrt{1-\frac{4m^{2}_{c}}{s}}}{2}$ and $b_{min}=\frac{am^{2}_{c}}{as-m^{2}_{c}}$. $e_{u}$, $e_{s}$ and $e_{c}$ are the electric charges of up, strange and charm quark, respectively, in the unit of the positron's electric charge.

\subsection{The spectral densities of the three-point correlation function}

We will give the explicit expressions of the three-point correlation functions in this subsection.

For $\Gamma^{1}_{(+)\mu\nu}(p,p_{1},p_{2})$, the spectral density is
\begin{equation}
\rho^{1}_{3(+)}(t_{1},t_{1},t_{2})=\rho^{1(0)}_{3(+)}(t_{1},t_{1},t_{2})+\rho^{1(3)}_{3(+)}(t_{1},t_{1},t_{2})
+\rho^{1(4)}_{3(+)}(t_{1},t_{1},t_{2})+\rho^{1(5)}_{3(+)}(t_{1},t_{1},t_{2}),
\end{equation}
where
\begin{equation}
\rho^{1(0)}_{3(+)}(t_{1},t_{1},t_{2})=\frac{1}{32\sqrt{2}\pi^{4}}\int^{a_{max}}_{a_{min}}da[3(a-1)at_{1}
+m^{2}_{c}+\frac{3m_{c}m_{s}}{4}]t_{2},
\end{equation}
\begin{equation}
\rho^{1(3)}_{3(+)}(t_{1},t_{1},t_{2})=-\frac{m_{s}\langle\bar{q}q\rangle}{8\sqrt{2}\pi^{2}}\int^{a_{max}}_{a_{min}}da[3(a-1)at_{1}
+m^{2}_{c}]\delta(t_{2}),
\end{equation}
\begin{eqnarray}
\rho^{1(4)}_{3(+)}(t_{1},t_{1},t_{2})=&&-\frac{m^{4}_{c}\langle g^{2}_{s}GG\rangle}{576\sqrt{2}\pi^{4}}\int^{1}_{0}da\frac{a-1}{a^{2}}\frac{d^{2}}{dt^{2}_{1}}\delta((a-1)at_{1}+m^{2}_{c})t_{2}
\nonumber\\&&+\frac{m^{2}_{c}\langle g^{2}_{s}GG\rangle}{384\sqrt{2}\pi^{4}}\int^{1}_{0}da\frac{a-1}{a}\frac{d}{dt_{1}}\delta((a-1)at_{1}+m^{2}_{c})t_{2}
\nonumber\\&&+\frac{m_{c}m_{s}\langle g^{2}_{s}GG\rangle}{512\sqrt{2}\pi^{4}}\int^{1}_{0}da\frac{(a-1)^{2}}{a}\frac{d}{dt_{1}}\delta((a-1)at_{1}+m^{2}_{c})t_{2}
\nonumber\\&&+\frac{m^{3}_{c}m_{s}\langle g^{2}_{s}GG\rangle}{1536\sqrt{2}\pi^{4}}\int^{1}_{0}da\frac{a-1}{a^{2}}\frac{d^{2}}{dt^{2}_{1}}\delta((a-1)at_{1}+m^{2}_{c})t_{2}
\nonumber\\&&+\frac{\langle g^{2}_{s}GG\rangle}{768\sqrt{2}\pi^{4}}\int^{1}_{0}da(a-1)a\delta((a-1)at_{1}+m^{2}_{c})t_{2}
\nonumber\\&&-\frac{m^{2}_{c}\langle g^{2}_{s}GG\rangle}{256\sqrt{2}\pi^{4}}\int^{1}_{0}da\frac{d}{dt_{1}}\delta((a-1)at_{1}+m^{2}_{c})t_{2}
\nonumber\\&&+\frac{m_{c}m_{s}\langle g^{2}_{s}GG\rangle}{1024\sqrt{2}\pi^{4}}\int^{1}_{0}da\frac{d}{dt_{1}}\delta((a-1)at_{1}+m^{2}_{c})t_{2}
\nonumber\\&&+\frac{\langle g_{s}GG\rangle}{384\sqrt{2}\pi^{4}}\int^{a_{max}}_{a_{min}}da(a-1)\nonumber\\&&-\frac{\langle g_{s}GG\rangle}{384\sqrt{2}\pi^{4}}\int^{a_{max}}_{a_{min}}da[3(a-1)at_{1}
+m^{2}_{c}]\delta(t_{2}),
\end{eqnarray}
\begin{eqnarray}
\rho^{1(5)}_{3(+)}(t_{1},t_{1},t_{2})=&&-\frac{m_{c}(\langle g_{s}\bar{q}\sigma Gq\rangle+\langle g_{s}\bar{s}\sigma Gs\rangle)}{48\sqrt{2}\pi^{2}}(a_{max}-a_{min})\delta(t_{2})\nonumber\\&&
+\frac{m_{s}\langle g_{s}\bar{s}\sigma Gs\rangle}{96\sqrt{2}\pi^{2}}\int^{a_{max}}_{a_{min}}da[3(a-1)at_{1}
+m^{2}_{c}]\frac{d}{dt_{2}}\delta(t_{2})\nonumber\\&&
+\frac{m^{2}_{c}m_{s}\langle g_{s}\bar{q}\sigma Gq\rangle}{12\sqrt{2}\pi^{2}}\frac{\delta(t_{2})}{\sqrt{t_{1}(t_{1}-4m^{2}_{c})}}\nonumber\\&&
-\frac{m_{s}\langle g_{s}\bar{q}\sigma Gq\rangle}{16\sqrt{2}\pi^{2}}\int^{a_{max}}_{a_{min}}da(a-1)\delta(t_{2}).
\end{eqnarray}

For $\Gamma^{1}_{(-)\mu\nu}(p,p_{1},p_{2})$, we find that the perturbative part and quark-condensation part contain only Lorentz structure $p_{1\nu}p_{2\mu}$. As a result, we choose this structure to obtain our sum rule for the strong decay form factor $g_{1(-)}$. The corresponding spectral density is
\begin{equation}
\rho^{1}_{3(-)}(t_{1},t_{1},t_{2})=\rho^{1(0)}_{3(-)}(t_{1},t_{1},t_{2})+\rho^{1(3)}_{3(-)}(t_{1},t_{1},t_{2})
+\rho^{1(4)}_{3(-)}(t_{1},t_{1},t_{2})+\rho^{1(5)}_{3(-)}(t_{1},t_{1},t_{2}),
\end{equation}
where
\begin{equation}
\rho^{1(0)}_{3(-)}(t_{1},t_{1},t_{2})=\frac{3m_{c}m_{s}}{64\sqrt{2}\pi^{4}}(a_{max}-a_{min}),
\end{equation}
\begin{equation}
\rho^{1(3)}_{3(-)}(t_{1},t_{1},t_{2})=\frac{m_{c}(\langle\bar{q}q\rangle-\langle\bar{s}s\rangle)}
{8\sqrt{2}\pi^{2}}(a_{max}-a_{min})\delta(t_{2}),
\end{equation}
\begin{eqnarray}
\rho^{1(4)}_{3(-)}(t_{1},t_{1},t_{2})=&&\frac{m_{c}m_{s}\langle g^{2}_{s}GG\rangle}{256\sqrt{2}\pi^{4}}\int^{1}_{0}da\frac{(a-1)^{2}}{a}\frac{d}{dt_{1}}\delta((a-1)at_{1}+m^{2}_{c})
\nonumber\\&&+\frac{m^{3}_{c}m_{s}\langle g^{2}_{s}GG\rangle}{768\sqrt{2}\pi^{4}}\int^{1}_{0}da\frac{a-1}{a^{2}}\frac{d^{2}}{dt^{2}_{1}}\delta((a-1)at_{1}+m^{2}_{c})
\nonumber\\&&+\frac{m_{c}m_{s}\langle g^{2}_{s}GG\rangle}{512\sqrt{2}\pi^{4}}\int^{1}_{0}da\frac{d}{dt_{1}}\delta((a-1)at_{1}+m^{2}_{c})
\nonumber\\&&+\frac{m_{c}m_{s}\langle g^{2}_{s}GG\rangle}{384\sqrt{2}\pi^{4}}\frac{\delta(t_{2})}{\sqrt{t_{1}(t_{1}-4m^{2}_{c})}},
\end{eqnarray}
\begin{equation}
\rho^{1(5)}_{3(-)}(t_{1},t_{1},t_{2})=\frac{m_{c}(\langle g_{s}\bar{q}\sigma Gq\rangle-\langle g_{s}\bar{s}\sigma Gs\rangle)}{16\sqrt{2}\pi^{2}}\frac{\delta(t_{2})}{\sqrt{t_{1}(t_{1}-4m^{2}_{c})}}.
\end{equation}
The spectral densities are proportional to either $m_{s}$, $\langle\bar{q}q\rangle-\langle\bar{s}s\rangle$ or $\langle g_{s}\bar{q}\sigma Gq\rangle-\langle g_{s}\bar{s}\sigma Gs\rangle$. As a result, the process $Z_{(-)cs}\rightarrow \eta_{c}K^{*}$ is suppressed in the present model.

For $\Gamma^{2}_{(+)\mu\nu}(p,p_{1},p_{2})$, we choose the Lorentz structure $g_{\mu\nu}$ to obtain the sum rule for the strong decay form factor $g_{2(+)}$. The corresponding spectral density is
\begin{equation}
\rho^{2}_{3(+)}(t_{1},t_{1},t_{2})=\rho^{2(0)}_{3(+)}(t_{1},t_{1},t_{2})+\rho^{2(3)}_{3(+)}(t_{1},t_{1},t_{2})
+\rho^{2(4)}_{3(+)}(t_{1},t_{1},t_{2})+\rho^{2(5)}_{3(+)}(t_{1},t_{1},t_{2}),
\end{equation}
where
\begin{equation}
\rho^{2(0)}_{3(+)}(t_{1},t_{1},t_{2})=\frac{3}{32\sqrt{2}\pi^{4}}\int^{a_{max}}_{a_{min}}da[(a-1)at_{1}
+\frac{m_{c}m_{s}}{4}]t_{2},
\end{equation}
\begin{equation}
\rho^{2(3)}_{3(+)}(t_{1},t_{1},t_{2})=-\frac{m_{s}(2\langle\bar{q}q\rangle-\langle\bar{s}s\rangle)}
{8\sqrt{2}\pi^{2}}\int^{a_{max}}_{a_{min}}da(a-1)at_{1}\delta(t_{2}),
\end{equation}
\begin{eqnarray}
\rho^{2(4)}_{3(+)}(t_{1},t_{1},t_{2})=&&-\frac{m^{4}_{c}\langle g^{2}_{s}GG\rangle}{384\sqrt{2}\pi^{4}}\int^{1}_{0}da\frac{a-1}{a^{2}}\frac{d^{2}}{dt^{2}_{1}}\delta((a-1)at_{1}+m^{2}_{c})t_{2}
\nonumber\\&&-\frac{m^{2}_{c}\langle g^{2}_{s}GG\rangle}{768\sqrt{2}\pi^{4}}\int^{1}_{0}da\frac{(a-1)^{2}}{a}\frac{d}{dt_{1}}\delta((a-1)at_{1}+m^{2}_{c})t_{2}
\nonumber\\&&+\frac{m^{2}_{c}\langle g^{2}_{s}GG\rangle}{256\sqrt{2}\pi^{4}}\int^{1}_{0}da\frac{a-1}{a}\frac{d}{dt_{1}}\delta((a-1)at_{1}+m^{2}_{c})t_{2}
\nonumber\\&&+\frac{m_{c}m_{s}\langle g^{2}_{s}GG\rangle}{512\sqrt{2}\pi^{4}}\int^{1}_{0}da\frac{(a-1)^{2}}{a}\frac{d}{dt_{1}}\delta((a-1)at_{1}+m^{2}_{c})t_{2}
\nonumber\\&&+\frac{m^{3}_{c}m_{s}\langle g^{2}_{s}GG\rangle}{1536\sqrt{2}\pi^{4}}\int^{1}_{0}da\frac{a-1}{a^{2}}\frac{d^{2}}{dt^{2}_{1}}\delta((a-1)at_{1}+m^{2}_{c})t_{2}
\nonumber\\&&+\frac{m^{2}_{c}\langle g^{2}_{s}GG\rangle}{1536\sqrt{2}\pi^{4}}\int^{1}_{0}da\frac{d}{dt_{1}}\delta((a-1)at_{1}+m^{2}_{c})t_{2}
\nonumber\\&&-\frac{m_{c}m_{s}\langle g^{2}_{s}GG\rangle}{3072\sqrt{2}\pi^{4}}\int^{1}_{0}da\frac{d}{dt_{1}}\delta((a-1)at_{1}+m^{2}_{c})t_{2}
\nonumber\\&&-\frac{\langle g^{2}_{s}GG\rangle}{768\sqrt{2}\pi^{4}}\int^{1}_{0}da(a-1)a\delta((a-1)at_{1}+m^{2}_{c})t_{2}
\nonumber\\&&+\frac{m^{2}_{c}\langle g^{2}_{s}GG\rangle}{96\sqrt{2}\pi^{4}}\int^{1}_{0}da(a-1)\delta((a-1)at_{1}+m^{2}_{c})
\nonumber\\&&+\frac{\langle g_{s}GG\rangle}{384\sqrt{2}\pi^{4}}\int^{a_{max}}_{a_{min}}da(a-1)\nonumber\\&&
+\frac{\langle g_{s}GG\rangle}{128\sqrt{2}\pi^{4}}\int^{a_{max}}_{a_{min}}da(a-1)at_{1}\delta(t_{2}),
\end{eqnarray}
\begin{equation}
\rho^{2(5)}_{3(+)}(t_{1},t_{1},t_{2})=\frac{m^{2}_{c}m_{s}\langle g_{s}\bar{q}\sigma Gq\rangle}{12\sqrt{2}\pi^{2}}\frac{\delta(t_{2})}{\sqrt{t_{1}(t_{1}-4m^{2}_{c})}}.
\end{equation}
In the above equations, $a_{max}=\frac{1+\sqrt{1-\frac{4m^{2}_{c}}{s}}}{2}$, and $a_{min}=\frac{1-\sqrt{1-\frac{4m^{2}_{c}}{s}}}{2}$.

For $\Gamma^{2}_{(-)\mu\nu}(p,p_{1},p_{2})$, we find that it is zero except the mixed condensation with Lorentz structure $g_{\mu\nu}$ and $p_{2\mu}p_{1\nu}$. Therefore, we believe that the process $Z_{(-)cs}\rightarrow J/\psi K$ is suppressed in the present model.

For $\Gamma^{3}_{(\pm)\mu\nu}(p,p_{1},p_{2})$, we find that their theoretical representations are the same. The spectral densities corresponding to the Lorentz structure $g_{\mu\nu}$ are
\begin{equation}
\rho^{3}_{3(\pm)}(t_{1},t_{1},t_{2})=\rho^{3(0)}_{3}(t_{1},t_{1},t_{2})+\rho^{3(3)}_{3}(t_{1},t_{1},t_{2})
+\rho^{3(4)}_{3}(t_{1},t_{1},t_{2})+\rho^{3(5)}_{3}(t_{1},t_{1},t_{2})+\rho^{3(6)}_{3}(t_{1},t_{1},t_{2}),
\end{equation}
where
\begin{equation}
\rho^{3(0)}_{3}(t_{1},t_{1},t_{2})=\frac{9}{16\sqrt{2}\pi^{4}}\int^{1}_{a_{min}}da[a(m^{2}_{c}-at_{2})-m_{c}m_{s}]
\int^{1}_{b_{min}}db(m^{2}_{c}-bt_{1}),
\end{equation}
\begin{eqnarray}
\rho^{3(3)}_{3}(t_{1},t_{1},t_{2})=&&\frac{3m_{c}\langle\bar{q}q\rangle}{4\sqrt{2}\pi^{2}}\int^{1}_{a_{min}}da[a(m^{2}_{c}-at_{2})-m_{c}m_{s}]
\delta(m^{2}_{c}-t_{1})\nonumber\\&&+\frac{3m_{c}\langle\bar{s}s\rangle}{4\sqrt{2}\pi^{2}}\int^{1}_{b_{min}}db(m^{2}_{c}-bt_{1})\delta(m^{2}_{c}-t_{2})
\nonumber\\&&-\frac{3m^{2}_{c}m_{s}\langle\bar{s}s\rangle}{8\sqrt{2}\pi^{2}}\int^{1}_{b_{min}}db(m^{2}_{c}-bt_{1})\frac{d}{dt_{2}}\delta(m^{2}_{c}-t_{2}),
\end{eqnarray}
\begin{eqnarray}
\rho^{3(4)}_{3}(t_{1},t_{1},t_{2})=&&\frac{m^{2}_{c}\langle g^{2}_{s}GG\rangle}{128\sqrt{2}\pi^{4}}\int^{1}_{a_{min}}da[a(m^{2}_{c}-at_{2})-m_{c}m_{s}]\int^{1}_{0}db\frac{1}{b}
\frac{d}{dt_{1}}\delta(m^{2}_{c}-bt_{1})\nonumber\\&&+\frac{\langle g^{2}_{s}GG\rangle}{128\sqrt{2}\pi^{4}}\int^{1}_{b_{min}}db(m^{2}_{c}-bt_{1})\int^{1}_{0}da
(m^{2}_{c}-\frac{3m_{c}m_{s}}{a})\frac{d}{dt_{2}}\delta(m^{2}_{c}-at_{2})\nonumber\\&&+\frac{m^{3}_{c}m_{s}\langle g^{2}_{s}GG\rangle}{128\sqrt{2}\pi^{4}}\int^{1}_{b_{min}}db(m^{2}_{c}-bt_{1})\int^{1}_{0}da
\frac{1}{a^{2}}\frac{d^{2}}{dt^{2}_{2}}\delta(m^{2}_{c}-at_{2})\nonumber\\&&-\frac{3\langle g^{2}_{s}GG\rangle}{128\sqrt{2}\pi^{4}}\int^{1}_{a_{min}}da[a(m^{2}_{c}-at_{2})-m_{c}m_{s}]\delta(m^{2}_{c}-t_{1})\nonumber\\&&
+\frac{\langle g^{2}_{s}GG\rangle}{128\sqrt{2}\pi^{4}}\int^{1}_{b_{min}}db(m^{2}_{c}-bt_{1})\delta(m^{2}_{c}-t_{2})\nonumber\\&&
+\frac{\langle g^{2}_{s}GG\rangle}{128\sqrt{2}\pi^{4}}\int^{1}_{b_{min}}db(m^{2}_{c}-bt_{1})\int^{1}_{0}daa\delta(m^{2}_{c}-at_{2}),
\end{eqnarray}
\begin{eqnarray}
\rho^{3(5)}_{3}(t_{1},t_{1},t_{2})=&&-\frac{3m^{3}_{c}\langle g_{s}\bar{q}\sigma Gq\rangle}{16\sqrt{2}\pi^{2}}\int^{1}_{a_{min}}da[a(m^{2}_{c}-at_{2})-m_{c}m_{s}]\frac{d^{2}}{dt^{2}_{1}}\delta(m^{2}_{c}-t_{1})
\nonumber\\&&-\frac{m^{2}_{c}(3m_{c}+m_{s})\langle g_{s}\bar{s}\sigma Gs\rangle}{16\sqrt{2}\pi^{2}}\int^{1}_{b_{min}}db(m^{2}_{c}-bt_{1})\frac{d^{2}}{dt^{2}_{2}}\delta(m^{2}_{c}-t_{2})
\nonumber\\&&+\frac{m^{4}_{c}m_{s}\langle g_{s}\bar{s}\sigma Gs\rangle}{16\sqrt{2}\pi^{2}}\int^{1}_{b_{min}}db(m^{2}_{c}-bt_{1})\frac{d^{3}}{dt^{3}_{2}}\delta(m^{2}_{c}-t_{2})
\nonumber\\&&+\frac{3m_{c}\langle g_{s}\bar{q}\sigma Gq\rangle}{8\sqrt{2}\pi^{2}}\int^{1}_{a_{min}}da[a(m^{2}_{c}-at_{2})-m_{c}m_{s}]\frac{d}{dt_{1}}\delta(m^{2}_{c}-t_{1}),
\end{eqnarray}
\begin{equation}
\rho^{3(6)}_{3}(t_{1},t_{1},t_{2})=\frac{m^{2}_{c}\langle\bar{q}q\rangle\langle\bar{s}s\rangle}{\sqrt{2}}
\delta(m^{2}_{c}-t_{1})\delta(m^{2}_{c}-t_{2})-\frac{m^{3}_{c}m_{s}\langle\bar{q}q\rangle\langle\bar{s}s\rangle}{2\sqrt{2}}
\delta(m^{2}_{c}-t_{1})\frac{d}{dt_{2}}\delta(m^{2}_{c}-t_{2}).
\end{equation}

For $\Gamma^{4}_{(\pm)\mu\nu}(p,p_{1},p_{2})$, we find that their theoretical representations are the same. The spectral densities corresponding to the Lorentz structure $g_{\mu\nu}$ are
\begin{equation}
\rho^{4}_{3(\pm)}(t_{1},t_{1},t_{2})=\rho^{4(0)}_{3}(t_{1},t_{1},t_{2})+\rho^{4(3)}_{3}(t_{1},t_{1},t_{2})
+\rho^{4(4)}_{3}(t_{1},t_{1},t_{2})+\rho^{4(5)}_{3}(t_{1},t_{1},t_{2})+\rho^{4(6)}_{3}(t_{1},t_{1},t_{2}),
\end{equation}
where
\begin{equation}
\rho^{4(0)}_{3}(t_{1},t_{1},t_{2})=\frac{9}{16\sqrt{2}\pi^{4}}\int^{1}_{a_{min}}da(m^{2}_{c}-m_{c}m_{s}-at_{2})
\int^{1}_{b_{min}}dbb(m^{2}_{c}-bt_{1}),
\end{equation}
\begin{eqnarray}
\rho^{4(3)}_{3}(t_{1},t_{1},t_{2})=&&\frac{3m_{c}\langle\bar{q}q\rangle}{4\sqrt{2}\pi^{2}}\int^{1}_{a_{min}}da(m^{2}_{c}-m_{c}m_{s}-at_{2})
\delta(m^{2}_{c}-t_{1})\nonumber\\&&+\frac{3(2m_{c}-m_{s})\langle\bar{s}s\rangle}{8\sqrt{2}\pi^{2}}\int^{1}_{b_{min}}dbb(m^{2}_{c}-bt_{1})\delta(m^{2}_{c}-t_{2})
\nonumber\\&&-\frac{3m^{2}_{c}m_{s}\langle\bar{s}s\rangle}{8\sqrt{2}\pi^{2}}\int^{1}_{b_{min}}dbb(m^{2}_{c}-bt_{1})\frac{d}{dt_{2}}\delta(m^{2}_{c}-t_{2}),
\end{eqnarray}
\begin{eqnarray}
\rho^{4(4)}_{3}(t_{1},t_{1},t_{2})=&&\frac{m^{2}_{c}\langle g^{2}_{s}GG\rangle}{128\sqrt{2}\pi^{4}}\int^{1}_{a_{min}}da(m^{2}_{c}-m_{c}m_{s}-at_{2})\int^{1}_{0}db
\frac{d}{dt_{1}}\delta(m^{2}_{c}-bt_{1})\nonumber\\&&+\frac{m_{c}(m_{c}-3m_{s})\langle g^{2}_{s}GG\rangle}{128\sqrt{2}\pi^{4}}\int^{1}_{b_{min}}dbb(m^{2}_{c}-bt_{1})\int^{1}_{0}da
\frac{1}{a}\frac{d}{dt_{2}}\delta(m^{2}_{c}-at_{2})\nonumber\\&&+\frac{m^{3}_{c}m_{s}\langle g^{2}_{s}GG\rangle}{128\sqrt{2}\pi^{4}}\int^{1}_{b_{min}}dbb(m^{2}_{c}-bt_{1})\int^{1}_{0}da
\frac{1}{a^{2}}\frac{d^{2}}{dt^{2}_{2}}\delta(m^{2}_{c}-at_{2})\nonumber\\&&+\frac{\langle g^{2}_{s}GG\rangle}{128\sqrt{2}\pi^{4}}\int^{1}_{a_{min}}da(m^{2}_{c}-m_{c}m_{s}-at_{2})\delta(m^{2}_{c}-t_{1})\nonumber\\&&
+\frac{\langle g^{2}_{s}GG\rangle}{128\sqrt{2}\pi^{4}}\int^{1}_{a_{min}}da(m^{2}_{c}-m_{c}m_{s}-at_{2})\int^{1}_{0}dbb\delta(m^{2}_{c}-bt_{1})\nonumber\\&&
-\frac{3\langle g^{2}_{s}GG\rangle}{128\sqrt{2}\pi^{4}}\int^{1}_{b_{min}}dbb(m^{2}_{c}-bt_{1})\delta(m^{2}_{c}-t_{2}),
\end{eqnarray}
\begin{eqnarray}
\rho^{4(5)}_{3}(t_{1},t_{1},t_{2})=&&-\frac{3m^{3}_{c}\langle g_{s}\bar{q}\sigma Gq\rangle}{16\sqrt{2}\pi^{2}}\int^{1}_{a_{min}}da(m^{2}_{c}-m_{c}m_{s}-at_{2})\frac{d^{2}}{dt^{2}_{1}}\delta(m^{2}_{c}-t_{1})
\nonumber\\&&-\frac{3m^{3}_{c}\langle g_{s}\bar{s}\sigma Gs\rangle}{16\sqrt{2}\pi^{2}}\int^{1}_{b_{min}}dbb(m^{2}_{c}-bt_{1})\frac{d^{2}}{dt^{2}_{2}}\delta(m^{2}_{c}-t_{2})
\nonumber\\&&+\frac{m^{4}_{c}m_{s}\langle g_{s}\bar{s}\sigma Gs\rangle}{16\sqrt{2}\pi^{2}}\int^{1}_{b_{min}}dbb(m^{2}_{c}-bt_{1})\frac{d^{3}}{dt^{3}_{2}}\delta(m^{2}_{c}-t_{2})
\nonumber\\&&+\frac{3m_{c}\langle g_{s}\bar{q}\sigma Gq\rangle}{8\sqrt{2}\pi^{2}}\int^{1}_{b_{min}}dbb(m^{2}_{c}-bt_{1})\frac{d}{dt_{2}}\delta(m^{2}_{c}-t_{2}),
\end{eqnarray}
\begin{eqnarray}
\rho^{4(6)}_{3}(t_{1},t_{1},t_{2})=&&\frac{m^{2}_{c}\langle\bar{q}q\rangle\langle\bar{s}s\rangle}{\sqrt{2}}
\delta(m^{2}_{c}-t_{1})\delta(m^{2}_{c}-t_{2})-\frac{m_{c}m_{s}\langle\bar{q}q\rangle\langle\bar{s}s\rangle}{2\sqrt{2}}
\delta(m^{2}_{c}-t_{1})\delta(m^{2}_{c}-t_{2})\nonumber\\&&-\frac{m^{3}_{c}m_{s}\langle\bar{q}q\rangle\langle\bar{s}s\rangle}{2\sqrt{2}}
\delta(m^{2}_{c}-t_{1})\frac{d}{dt_{2}}\delta(m^{2}_{c}-t_{2}).
\end{eqnarray}
In the above equations, $a_{min}=m^{2}_{c}/t_{2}$ and $b_{min}=m^{2}_{c}/t_{1}$.
\end{appendix}

\bibliography{Ref}
\bibliographystyle{unsrt}

\end{document}